\documentclass[%
 reprint,
 unsortedaddress,
nofootinbib,
 amsmath,amssymb,
 aps,
 pra,
 floatfix,
]{revtex4-2}

\usepackage{graphicx}
\usepackage{dcolumn}
\usepackage{bm}

\usepackage{lipsum}
\usepackage{comment}
\usepackage{color}
\usepackage{braket}

\newcommand{\murm}{%
  \ifmmode
    \mathchoice
        {\hbox{\normalsize\textmu}}
        {\hbox{\normalsize\textmu}}
        {\hbox{\scriptsize\textmu}}
        {\hbox{\tiny\textmu}}%
  \else
    \textmu
  \fi
}

\newcommand{\ddm}{\mathrm{dd}\murm} 
\newcommand{\ddmr}{\mathrm{dd}\murm^\ast} 
\newcommand{\dtm}{\mathrm{dt}\murm} 
\newcommand{\ttm}{\mathrm{tt}\murm} 
\newcommand{\dtmr}{\mathrm{dt}\murm^\ast} 
 
\newcommand{\dm}{\mathrm{d}\murm} 
\newcommand{\mcf}{$\murm$CF} 
\newcommand{\add}[1]{#1}

\graphicspath{{}}

\begin{document}

\def\textfraction{0.01}

\preprint{APS/123-QED}

\title{%
Radiative decay of muonic molecules in resonance states
}

\author{Takuma Yamashita}
\email{tyamashita@tohoku.ac.jp}
\affiliation{Institute for Excellence in Higher Education, Tohoku University, Sendai 980-8576, Japan}
\affiliation{Department of Chemistry, Tohoku University, Sendai, 980-8578, Japan}

\author{Kazuhiro Yasuda}%
\author{Yasushi Kino}%
\affiliation{Department of Chemistry, Tohoku University, Sendai, 980-8578, Japan}

\date{\today}

\begin{abstract}
In this study, we theoretically investigated x-ray spectra from the radiative decay of muonic deuterium molecules in resonance states dd$\murm^\ast$,
which plays an important role in a new kinetic model of muon catalyzed fusion ($\murm$CF).
The resonance states are Feshbach resonances located below the d$\murm$($n=2$) + d threshold energy and radiatively decay into the continuum or bound states. 
The x-ray spectra having characteristic shapes according to the radial distribution of the two nuclei
are obtained using precise three-body wave functions.
We carefully examined the convergence of the x-ray spectra and achieved agreements between the length- and velocity-gauge calculations.
We revealed a non-adiabatic kinetic energy distribution of the decay fragments, indicating that 
the radiative decay becomes a heating source of muonic atoms.
We also investigated the decay branch that directly results in bound-state muonic molecules.
Some resonance states dd$\murm^\ast$ and dt$\murm^\ast$ are found to have high branching ratios to the bound state
where intramolecular nuclear fusion occurs.
The formation of the muonic molecules in the bound states from metastable muonic atoms 
can be a fast track in the $\murm$CF cycle which skips a slow path to form the bound state from the ground-state muonic atoms
and increases the $\murm$CF cycle rate.
Although the spectra from the radiative decays are located in the energy range of $1.5$--$1.997$ keV, which is
close to the K$\alpha$ x-ray of 1.997 keV from muonic deuterium atoms, 
state-of-the-art microcalorimeters can distinguish them.
\end{abstract}

\maketitle

\section{Introduction}
\label{intro}

A muon ($\murm$), an elementary particle with a mass 207 times greater than that of an electron, forms a compact molecule with two hydrogen nuclei. 
When the nuclei are deuteron or triton, nuclear fusion occurs in the muonic hydrogen molecules, i.e., $\ddm$, $\dtm$, and $\ttm$.
The muon released after the fusion repeatedly causes the nuclear fusion reactions in hydrogen targets during its lifetime of 2.2 $\murm$s, 
which is so-called muon catalyzed fusion ($\murm$CF)~\cite{Breunlich1989,Ponomarev1990,Rafelski_1991,Froelich1992},
and the cyclic process is called $\murm$CF cycle.
For applications of the $\murm$CF, it is necessary to increase the cycle rate by elucidating the elementary processes of the muonic atoms in the hydrogen.
We recently proposed a new kinetic model of $\murm$CF~\cite{Yamashita:2022vi} which contained muonic molecules in resonance states 
and well reproduced the experimental $\murm$CF cycle rates in D$_2$-T$_2$ mixture.

The muonic hydrogen atom is initially formed in a highly excited state ($n\approx14$) and subsequently cascades to lower levels~\cite{Cohen_2004}. 
One of the crucial problems in cascade processes is the dynamics involving the metastable state of the muonic hydrogen atom, i.e., the $2s$ state, 
which has been investigated theoretically~\cite{Froelich1995,Wallenius1996,Kino1996,Pomerantsev2006,Popov2011,Popov2022,Yamashita:2022vi} 
and experimentally by laser spectroscopy~\cite{Pohl2001cw,Pohl2006,Ludhova2007,Diepold2013}.

The metastable state of a muonic deuterium atom d$\murm(2s)$ can produce a muonic deuterium molecule in the resonance states as follows.
\begin{align}
\mathrm{d}\murm(2s) + \mathrm{D}_2(\mathcal{L}_\mathrm{i},v_\mathrm{i}) \to [(\ddmr)\mathrm{dee}](\mathcal{L}_\mathrm{f},v_\mathrm{f}),
\label{vesman2}
\end{align}
where $\mathcal{L}_\mathrm{i/f},v_\mathrm{i/f}$ are the rotational and vibrational quantum numbers of the initial/final states, respectively.
$\ddmr$ denotes a muonic deuterium molecule $\ddm$ in a resonance state, in which the muonic molecular orbital is excited.
The final-state molecule $[(\ddmr)\mathrm{dee}]$ is a complex system in which $\ddmr$ behaves as a quasi-nucleus, and the two electrons bind 
the two nuclei $\ddmr$ and d.
The reaction~(\ref{vesman2}) was proposed as an analogy of the $\ddm$ formation mechanism, called a Vesman mechanism~\cite{Vesman1967}, which is presented as follows:
\begin{align}
\mathrm{d}\murm(1s) + \mathrm{D}_2(\mathcal{L}_\mathrm{i},v_\mathrm{i}) \to [(\ddm)\mathrm{dee}](\mathcal{L}_\mathrm{f},v_\mathrm{f}).
\label{vesman}
\end{align}
The excess energy, i.e., the sum of the binding energy of dd$\murm$ ($\sim1.97$ eV) and collision energy, is resonantly transferred 
to the rovibrational excitation energy of the two nuclei in the $[(\ddm)\mathrm{dee}]$ molecule. 
Because the resonance states of the muonic molecules $\ddmr$ are supported by long-range ion-dipole interactions between an excited muonic atom and the \add{nucleus~\cite{Shimamura1989,Hara1989,Wallenius1996b,Kar2007,Kar2008a}}, 
the energy levels of $\ddmr$ accumulate below the dissociation limit, thus providing a denser level density and more suitable conditions for the Vesman mechanism.
Therefore, the formation rate of the muonic molecules in the resonance states is estimated to be significantly higher than 
that of $\ddm$~\cite{Froelich1995,Wallenius1996,Kino1996}.

The fate of the $2s$ state of the muonic atom has involved a puzzle.
The quenching rate of the $2s$ state of the muonic atom (p$\murm$ in the H$_2$ target and $\dm$ in the D$_2$ target) 
is smaller than that predicted by a conventional cascade model~\cite{Ludhova2007}.
As the formation of the muonic molecules in the resonance states quenches the $2s$ state of the muonic atom, 
the reaction~(\ref{vesman2}) can be a mechanism to explain the observed population of the $2s$ state of the muonic atoms.
Furthermore, deexcitation processes via $\dtmr$ in the D$_2$-T$_2$ mixture target 
increase d$\mu(1s)$ population and
can explain the observed \mcf\ cycle rate~\cite{Froelich1995,Kino1996,Yamashita:2022vi}.
However, a cascade model has recently been developed~\cite{Pomerantsev2006,Popov2011,Popov2022} 
that indicates the importance of the direct Coulomb decay and collision-induced radiative quenching, 
which explains the observed population of p$\murm(2s)$ and $\dm(2s)$ without assuming the reaction~(\ref{vesman2}).

In this study, we calculate the x-ray spectra of the radiative decay of the resonance states $\ddmr$ into the continuum state,
\begin{align}
\ddmr \to \dm(1s) + \mathrm{d} + \gamma,
\label{raddis}
\end{align}
or into a bound state,
\begin{align}
\ddmr \to \mathrm{dd\murm} + \gamma.
\label{radbnd}
\end{align}
The process~(\ref{raddis}) produces x-rays with characteristic energy profiles 
depending on their rotational and vibrational states~\cite{Lindroth2003,Kilic2004}
in the energy range of $1.5$--$1.997$ keV.
The process~(\ref{radbnd}) produces, in contrast, monoenergetic x-rays whose energy differs from the $K_\alpha$ x-rays of d$\murm$ (1.997 keV).
The x-ray spectra, therefore, can be footprints demonstrating the presence of these molecules.
Thus far, it has been difficult to distinguish the x-rays of the process~(\ref{raddis}) having a broad spectrum from mono-energetic $K_\alpha$ x-rays
because of the energy resolution of the detectors.
Recently, high-resolution x-ray spectroscopy with microcalorimeters has been successfully applied to determine the energy levels of
muonic atoms~\cite{Okada2020,Okumura2021,Okumura2023} and show sufficient performance for x-ray energy of several keV with FWHM of 6 eV.
The x-ray spectra from $\ddmr$ obtained via the process~(\ref{raddis}) were theoretically reported 
in previous studies~\cite{Lindroth2003,Kilic2004} for several states.

To analyze the forthcoming experiments, as $\ddmr$ is expected to form in various rotational and vibrational states,
more comprehensive study on the x-ray spectra is required. Furthermore, the x-ray intensity from process~(\ref{radbnd})
has not been previously investigated while the process~(\ref{radbnd}) could introduce a direct pathway from metastable muonic atoms to the bound states of muonic molecules 
and play an important role in the \mcf\ cycle.
Thus, this study investigates the radiative decay of muonic molecules in the resonance states 
and predicts the x-ray spectra from the rotational states of $J=0$--$3$ and vibrational states of $\upsilon=0$--$8$.
 
We solve the Schr\"{o}dinger equation for a three-body system using the Gaussian expansion method~\cite{GEM2003} 
after separating the center-of-mass motion. 
The transition rates are calculated using the complex scaling method~\cite{Ho1983,Buchleitner1994} under the dipole approximation. 
We first examine the accuracy of x-ray spectra, and investigate their characteristic shape associated with radial distribution functions of the resonance states. 
We also reveal the radiative decay mechanism from the point of view of kinetic energy distributions of the decay fragments in comparison with adiabatic approximations
because the kinetic energy distribution of the decay fragments is important to evaluate epi-thermal muonic atoms~\cite{Cohen1986,Markushin1993,Fujiwara2000}. 
Finally we discuss the competition between the radiative decays (\ref{raddis}) and (\ref{radbnd}).

The remainder of this paper is organized as follows. Sec.~\ref{theory} outlines the theoretical calculations. 
Sec.~\ref{RandD} presents the x-ray spectra of each resonance state, 
angular momentum dependency of the x-ray spectra, and radiative decay rate into the bound states.
Sec.~\ref{conclusion} summarizes the discussion. 
Atomic units (a.u.; $m_{\rm e}=\hbar=e=1$) and muonic atomic units (m.a.u.; $m_{\murm}=\hbar=e=1$) are used throughout this paper, except when specified otherwise.

\section{Theory}
\label{theory}

\subsection{Complex coordinate rotation method for three-body systems}
 
\begin{figure}[t]
\begin{center}
\resizebox{0.45\textwidth}{!}{%
\includegraphics[bb=0 0 530 238]{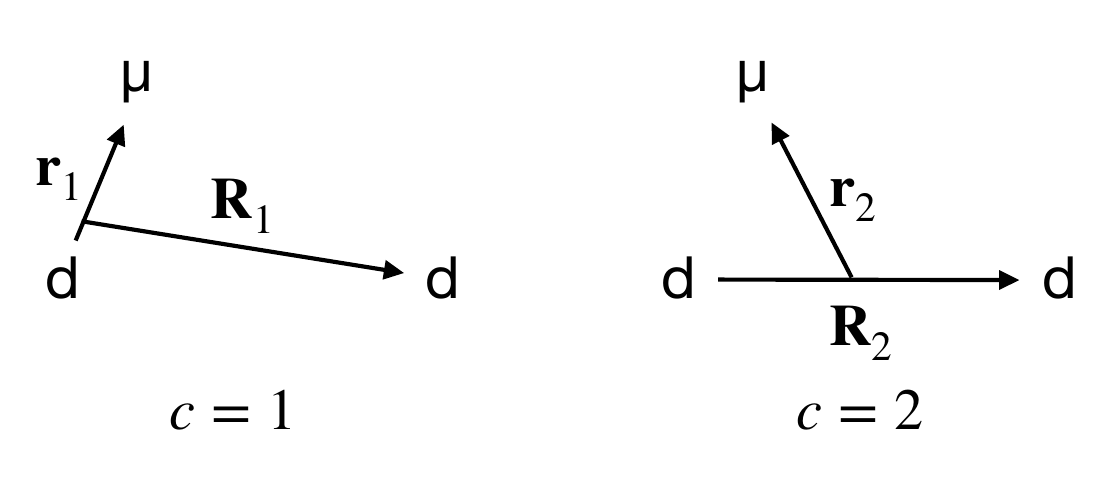}}
\caption{Jacobi coordinate systems used in this study.}
\label{f_coord}
\end{center}
\end{figure}

We consider the three-body Hamiltonian 
involving kinetic energy operators (separated from the motion of the center of mass)
and the Coulomb potential energy operators:
\begin{equation}
 \label{Hamil01}
  \hat H
    = \sum_{i=1}^{3} -\frac{1}{2m_i} \nabla_{r_i}^2 
      + \frac{1}{2m_{\rm G}} \nabla_{r_{\rm G}}^2    
     + \frac{1}{r_{12}}
      - \frac{1}{r_{13}}
      - \frac{1}{r_{23}},
\end{equation}
where $m_i$ is the mass of particle $i$ ($i=1$: d, $i=2$: d, and $i=3$: $\murm$), 
$m_{\rm G}$ is the mass of the system, $\nabla_{r_i}$ is a differential operator for the position vector 
${\bf r}_i$, $\nabla_{r_{\rm G}}$ is a differential operator for the center of mass, 
and $r_{ij}$ is the relative distance between the particles $i$ and $j$.
We use the deuteron-muon mass ratio $m_\mathrm{d}/m_{\murm}=17.751\,674\,54$~\cite{CODATA1998} for comparison with the previous study~\cite{Kilic2004}.
The muon-electron mass ratio $m_{\murm}/m_\mathrm{e}=206.768\,283$~\cite{CODATA2018} is used throughout this work.

We adopt a complex coordinate rotation (CCR) method~\cite{Ho1983} to calculate the resonance states and their radiative decays.
A CCR Hamiltonian $H(\theta)$ is formally written as
\begin{align}
H(\theta) = R(\theta) H R(-\theta),
\end{align}
where $R(\theta)$ denotes the complex rotation operator.
$H(\theta)$ corresponds to the Hamiltonian where a complex factor of $e^{\mathrm{i}\theta}$ scales the distance $r$ as $r\rightarrow re^{\mathrm{i}\theta}$.

The eigen functions $\psi_{J,v}(\theta)$ of $H(\theta)$ are expanded in terms of $L^2$ integrable basis functions 
and obtained by solving the generalized complex eigenvalue problem,
\begin{align}
\langle  \bar{\psi}_{J,v'}(\theta) | H(\theta) | \psi_{J,v}(\theta) \rangle &= E_{J,v}(\theta) \delta_{v'v},
\label{e_complex_eigen}
\end{align}
where 
$J$ is the total orbital angular momentum quantum number, 
$v=1,\cdots$ denotes the index of the eigenstates,
and
$\bar{\psi}_{J,v}(\theta)$ denotes the complex conjugate of $\psi_{J,v}(\theta)$, except for the angular part.
For resonance states, the complex eigenvalues $E_{J,v}(\theta)$ obtained by the CCR method take stationary values at $\theta=\theta'$ on the complex energy plane, 
with the real part for the resonance energy $E_{J,v}^{\mathrm{(R)}}$ and the imaginary part for the resonance width $\varGamma$: 
\begin{align}
E_{J,v}(\theta') = E_{J,v}^{\mathrm{(R)}} -\mathrm{i}\varGamma/2.
\label{resolevel}
\end{align}
For continuum states, the eigenvalues exist on a line rotated by $-2\theta$ from the real axis, centered around the threshold energy $E_\mathrm{th}$ as
\begin{align}
\label{rel_kinetic_rot}
E_{J,v}(\theta) \simeq E_\mathrm{th} + K_\mathrm{rel} (\cos 2\theta - \mathrm{i} \sin 2\theta),
\end{align}
where $K_\mathrm{rel}$ denotes the kinetic energy of the relative motion of the fragments.

The wave function $\psi_{J,v}(\theta)$ is expressed as
\begin{align}
\label{wf1}
&\psi_{J,v}(\theta) = \sum_{c=1}^{2} \sum_{l_{ci},L_{cj}} \sum_{ij} (1\pm P_\mathrm{dd}) \nonumber \\
&\times \left\{ C_{cijl_{ci}L_{cj}}^{(v)}(\theta) \, r_c^{l_{ci}}R_c^{L_{cj}} 
                \exp\left(-a_i r_c^2 -A_j R_c^2\right) \sin(\beta A_j R_c^2) \right. \nonumber\\
& \left. + D_{cijl_{ci}L_{cj}}^{(v)}(\theta) \, r_c^{l_{ci}}R_c^{L_{cj}} 
                \exp\left(-a_i r_c^2 -A_j R_c^2\right) \cos(\beta A_j R_c^2) \right\}
\nonumber \\ 
& \times \left[ Y_{l_{ci}}(\hat{\mathbf{r}}_c) \otimes Y_{L_{cj}}(\hat{\mathbf{R}}_c)\right]_{JM},
\end{align}
where $M$ denotes a projection of the total angular momentum $J$ onto the $z$-axis.
$c$ denotes the coordinate system specified by $\{\mathbf{r}_c,\mathbf{R}_c\}$ 
(Fig.~\ref{f_coord}); $P_\mathrm{dd}$ is a permutation operator for two identical deuterons; 
$a_i$, $A_j$, and $\beta$ are real numbers; and 
$Y_{l_{ci}(L_{cj})}$ denotes the spherical harmonics of angular momentum quantum number $l_{ci}(L_{cj})$.
$\left[\cdots \right]$ denotes 
a tensor product of two angular momentum states given by the linear combination of spherical harmonics 
with Clebsch--Gordan coefficients as defined in Ref.~\cite{Varshalovich_QTAM}.
The $\beta\neq0$ introduces oscillating Gaussian functions~\cite{GEM2003}. These basis functions are suitable for describing the 
vibrationally excited state, the inter-nuclear wave function of which increases the number of nodes.
In this study, we set $\beta=1.5$. 

As the deuteron has spin 1, the total wave function, including the spin part, must be symmetric against the permutation of the two d nuclei.
When the two d nuclei configure the total nuclear spin $S_\mathrm{nucl}=0$ or $2$, the spin part is symmetric; 
however, in the $S_\mathrm{nucl}=1$ case, 
the spin part is antisymmetric against the permutation of the two deuterons. 
Thus, we select $(1+P_\mathrm{dd})$ for $S_\mathrm{nucl}=0$ and $2$ and $(1-P_\mathrm{dd})$ for $S_\mathrm{nucl}=1$.
The spatial parity of the wave function is determined by $\Pi=(-1)^{l_{ci}+L_{cj}}$. 
Hereafter, we distinguish the symmetry of the $\ddm$ states by the total nuclear spin $S_\mathrm{nucl}$, total orbital angular momentum quantum number $J$,
and the spatial parity $\Pi$ (\textit{even} or \textit{odd}) and denote the symmetry as $^{2S_\mathrm{nucl}+1}J^{e/o}$. 

The convergence with respect to the number of angular momenta $l_{ci}$ and $L_{cj}$ when using the two coordinate systems is much faster than
the convergence when using a single coordinate system.
The basis functions written in $c=1$ and $c=2$ are suited for the description of the d$\murm$-d interaction
and muonic molecular orbital around two deuterons, respectively.
The linear coefficients $C_{cijl_{ci}L_{cj}}^{(v)}(\theta)$ and $D_{cijl_{ci}L_{cj}}^{(v)}(\theta)$ which are 
complex values depending on $\theta$ are determined by Eq.~(\ref{e_complex_eigen}).
Hereafter, we denote the total number of basis functions, namely the number of linear coefficients, by $N_\mathrm{max}$.

\subsection{Calculation of resonance-continuum x-ray spectrum}

We calculate the radiative decay rates of the $\ddmr$ in the rovibrational state $(J,v_\mathrm{r})$ where $v_\mathrm{r}$ denotes the vibrational quantum number.
The energy derivative of the radiative decay rate of the resonance state into a continuum state, 
$\mathrm{d}\varGamma_\mathrm{RC}/\mathrm{d}E_\gamma$, can be calculated as a function of x-ray energy $E_\gamma$ by dipole approximation 
\begin{equation}
\frac{\mathrm{d}\varGamma_\mathrm{RC}}{\mathrm{d}E_\gamma} = \frac{4}{3}\alpha^3 E_\gamma^3 
\big| \langle \Psi_\mathrm{C}(E_\mathrm{f})  | \mathbf{d} |  \Psi_\mathrm{R} \rangle \big|^2,
\label{rate}
\end{equation}
where 
$\Psi_\mathrm{R}$ is a wave function of the resonance state, and
$\Psi_\mathrm{C}$ is an energy-normalized continuum state wave function corresponding to the energy $E_\mathrm{f}$.
$E_\gamma$ denotes the x-ray energy, $\alpha$ is the fine structure constant, and
$\mathbf{d}$ denotes the electric dipole moment operator.
The energy of the continuum state $E_\mathrm{f}$ satisfies 
\begin{align}
E_\mathrm{f}=E_{J,v_\mathrm{r}}^{\mathrm{(R)}}-E_\gamma,
\end{align}
where $E_{J,v_\mathrm{r}}^{\mathrm{(R)}}$ is the resonance energy of $\ddmr$.

The $\mathrm{d}\varGamma_\mathrm{RC}/\mathrm{d}E_\gamma$ are numerically calculated 
by the complex coordinate rotation method~\cite{Rescigno1975,Ho1983,Buchleitner1994,Kilic2004}.
The energy-normalized continuum state wave function satisfies
\begin{align}
\label{green1}
\ket{\Psi_\mathrm{C}(E_\mathrm{f})}\bra{\bar{\Psi}_\mathrm{C}(E_\mathrm{f})} = \frac{1}{2\mathrm{i}\pi}(G^-(E_\mathrm{f})-G^+(E_\mathrm{f})).
\end{align}
The $G^\pm(E_\mathrm{f})$ are the Green functions of the Hamiltonian on the real axis as
\begin{align}
G^\pm(E_\mathrm{f}) = \frac{1}{E_\mathrm{f}\pm \mathrm{i}\epsilon -H},
\end{align}
where $\epsilon$ is a small positive number to avoid singularity.
The Green function of the complex-rotated Hamiltonian $H(\theta)$ is related to $G^\pm(E_\mathrm{f})$ as
\begin{align}
\label{green_complex}
G^\pm(E_\mathrm{f}) = R(\mp\theta)\frac{1}{E_\mathrm{f}-H(\pm\theta)} R(\pm\theta),
\end{align}
for $\theta>0$.
Equation~(\ref{green1}) can be rewritten by Eq.~(\ref{green_complex}) as
\begin{align}
\label{green2}
&\ket{\Psi_\mathrm{C}(E_\mathrm{f})}\bra{\bar{\Psi}_\mathrm{C}(E_\mathrm{f})}  \nonumber\\
&= \frac{1}{2\mathrm{i}\pi}
\left[
 R(\theta)\frac{1}{E_\mathrm{f} - H(-\theta)} R(-\theta) - R(-\theta)\frac{1}{E_\mathrm{f} - H(\theta)} R(\theta)
\right].
\end{align}

We consider the eigen functions $\{\psi_{J_\mathrm{f},v}(\theta)\}$ where $J_\mathrm{f}$ is 
the total angular momentum quantum number of the whole system of the decay fragments. 
Under the dipole approximation, $J_\mathrm{f}=J\pm1$.  
Because the complex-rotated wave functions satisfy the following closure relation in a finite region of space~\cite{GEM2003}, 
\begin{align}
\label{closure}
\sum_{v} \ket{\psi_{J_\mathrm{f},v}(\theta)}\bra{\bar{\psi}_{J_\mathrm{f},v}(\theta)} =1,
\end{align}
Eq.~(\ref{green2}) becomes
\begin{align}
\label{green3}
&\ket{\Psi_\mathrm{C}(E_\mathrm{f})}\bra{\bar{\Psi}_\mathrm{C}(E_\mathrm{f})} 
= \frac{1}{2\mathrm{i}\pi}
\left[
 \frac{R(\theta)\ket{\psi_{J_\mathrm{f},v}(\theta)}\bra{\bar{\psi}_{J_\mathrm{f},v}(\theta)}R(-\theta)}{E_\mathrm{f} - \bar{E}_{J_\mathrm{f},v}(\theta)} \right. \nonumber \\
& \left.-\frac{R(-\theta)\ket{\psi_{J_\mathrm{f},v}(\theta)}\bra{\bar{\psi}_{J_\mathrm{f},v}(\theta)}R(\theta)}{E_\mathrm{f} - {E}_{J_\mathrm{f},v}(\theta)}
\right].
\end{align}
We note that the set of eigen functions obtained by the Gaussian expansion method becomes approximately a complete set in a finite region~\cite{GEM2003}.
Using Eq.~(\ref{green3}), we have the $\mathrm{d}\varGamma_\mathrm{RC}/\mathrm{d}E_\gamma$
in terms of $\{\psi_{J_\mathrm{f},v}(\theta)\}$ as in ~\cite{Buchleitner1994,Kilic2004,Yamashita2022c}
\begin{align}
\frac{\mathrm{d}\varGamma_\mathrm{RC}}{\mathrm{d}E_\gamma}
= \frac{4}{3}\alpha^3 E_\gamma^3 \frac{1}{\pi} \mathrm{Im} \sum_{v=1}^{v_\mathrm{max}} \left[
\frac{\langle \bar{\psi}_{J_\mathrm{f},v}(\theta) | \mathbf{d}(\theta) | \Psi_\mathrm{R}(\theta) \rangle^2}{E_{J_\mathrm{f},v}(\theta)-E_\mathrm{f}}
 \right],
\label{e_raddec}
\end{align}
where $\mathbf{d}(\theta)$ is the complex-rotated electric dipole moment operator.
The $\Psi_\mathrm{R}(\theta)$ is the rotated wave function of the resonance state which is a member of $\{\psi_{J,v}(\theta)\}$.
For the present system, the wave function of the resonance state $\Psi_\mathrm{R}$ has a small scattering component, and 
the complex-rotated wave function, $\Psi_\mathrm{R}(\theta)$, can be expanded in terms of $L^2$  
basis functions with a sufficient accuracy.
The number of eigen functions $v_\mathrm{max}$ is less than the total number of basis functions $N_\mathrm{max}$, namely $v_\mathrm{max}\leq N_\mathrm{max}$.
We will see in the following section that ${\mathrm{d}\varGamma_\mathrm{RC}}/{\mathrm{d}E_\gamma}$ converges as increasing $v_\mathrm{max}$.
Typically, we use $v_\mathrm{max}\sim400$, which is much smaller than $N_\mathrm{max}\sim10^4$.

To estimate the accuracy of our calculations, we calculate $\mathrm{d}\varGamma_\mathrm{RC}/\mathrm{d}E_\gamma$ 
in both length and velocity gauges.
Since the accuracy of the long-range component of the wave function affects the length-gauge calculation than the velocity-gauge one, 
the latter is better than the former because the complex coordinate rotation method artificially dampens the outgoing component.

\section{Results and Discussion}
\label{RandD}

\subsection{Energy levels of resonance states}

\begin{table}[t]                
\begin{ruledtabular}                
\caption[short table name.]{
Comparison of resonance energies calculated in this study with those of previous studies.
These values are given in eV relative to the d$\murm(n=2)$ + d threshold energy. 
\add{Resonance widths of these resonance states are estimated to be smaller than the significant values of this work (less than 10 $\mu$eV).}
}
\label{tbl_ddmueng_comp}
\begin{tabular}{ccrrrr}
Symmetry    &  $v_\mathrm{r}$ &  This work
            & Ref.~\cite{Kilic2004}   &  Ref.~\cite{Lindroth2003} & Ref.~\cite{Hara1989} \\ \hline
$^{1,5}S^e$ & $ 0$  & $218.111\,1$ & $218.111\,567$ & $218.112$   & $218.113$  \\  
$^{1,5}S^e$ & $ 1$  & $135.278\,5$ & $135.279\,003$ & $135.279$   & $135.278$  \\  
$^{1,5}S^e$ & $ 2$  & $ 72.966\,2$ &  $72.967\,058$ &  $72.697$   &  $72.962$  \\  
$^{1,5}S^e$ & $ 3$  & $ 31.901\,1$ &  $31.901\,769$ &  $31.902$   &  $31.884$  \\  
$^{1,5}S^e$ & $ 4$  & $ 12.616\,5$ &  $12.616\,688$ &  $12.617$   &  $12.606$  \\
$^{1,5}S^e$ & $ 5$  & $  5.311\,2$ &   $5.311\,346$ &   $5.311$   &   $5.304$  \\
$^{1,5}S^e$ & $ 6$  & $  2.275\,0$ &   $2.275\,273$ &             &   $2.210$  \\
$^{1,5}S^e$ & $ 7$  & $  0.981\,0$ &   $0.981\,232$ &             &            \\
$^{1,5}S^e$ & $ 8$  & $  0.424\,1$ &                &             &            \\
\hline
$^{  3}S^e$ & $ 0$ & $ 21.1551$ &                   &             & $21.156$    \\
$^{  3}S^e$ & $ 1$ & $  9.4149$ &                   &             &  $9.415$    \\
$^{  3}S^e$ & $ 2$ & $  4.0801$ &                   &             &  $4.080$   \\
$^{  3}S^e$ & $ 3$ & $  1.7656$ &                   &             &  $1.603$    \\
$^{  3}S^e$ & $ 4$ & $  0.7645$ &                   &             &             \\
$^{  3}S^e$ & $ 5$ & $  0.3311$ &                   &             &             \\
\hline
$^{  3}P^o$ & $ 0$  & $211.9236$ &                  &             & $211.926$    \\  
$^{  3}P^o$ & $ 1$  & $130.3486$ &                  &             & $130.348$    \\  
$^{  3}P^o$ & $ 2$  & $ 69.2351$ &                  &             & $69.225$     \\  
$^{  3}P^o$ & $ 3$  & $ 29.5255$ &                  &             & $29.504$     \\ 
$^{  3}P^o$ & $ 4$  & $ 11.4945$ &                  &             & $11.478$       \\ 
$^{  3}P^o$ & $ 5$  & $  4.7732$ &                  &             &  $4.758$       \\ 
$^{  3}P^o$ & $ 6$  & $  2.0157$ &                  &             &  $1.913$       \\ 
$^{  3}P^o$ & $ 7$  & $  0.8567$ &                  &             &                \\ 
$^{  3}P^o$ & $ 8$  & $  0.3650$ &                  &             &                \\ 
\hline 
$^{1,5}P^o$ & $ 0^\ast$  & $ 22.6458$ &             &             & $22.648$        \\ 
$^{1,5}P^o$ & $ 0$       & $ 20.1211$ &             &             & $20.122$        \\
$^{1,5}P^o$ & $ 1$       & $  8.8046$ &             &             & $8.805$         \\
$^{1,5}P^o$ & $ 2$       & $  3.7575$ &             &             & $3.749$         \\
$^{1,5}P^o$ & $ 3$       & $  1.6023$ &             &             & $1.395$         \\
$^{1,5}P^o$ & $ 4$       & $  0.6837$ &             &             &                 \\
$^{1,5}P^o$ & $ 5$       & $  0.2918$ &             &             &               
\end{tabular}
\end{ruledtabular}
\end{table}

To examine the accuracy of the resonance state wave functions, 
we investigated the dd$\murm^\ast$ resonance energies of $S$, $P$, $D$, and $F$ waves using a stabilization method~\cite{Hazi1970,Simons1981,Mandelshtam1993}.
The resonance energy levels of the $S$ and $P$ waves obtained in this study are listed in Table~\ref{tbl_ddmueng_comp} 
with the vibrational quantum number $v_\mathrm{r}$
and compared with those of several previous studies ~\cite{Hara1989,Lindroth2003,Kilic2004}.
Table~\ref{tbl_ddmueng_comp} expresses the resonance energies by quasi-binding energies,
\begin{align}
\varepsilon_{J,v_\mathrm{r}} = E_\mathrm{th}^{(n=2)} - E_{J,v_\mathrm{r}}^{\mathrm{(R)}},
\end{align} 
where $E_\mathrm{th}^{(n=2)}$ denotes the d$\murm(n=2)$ + d threshold energy.
Our calculations agree well with the latest complex coordinate rotation calculations ~\cite{Kilic2004} for $^{1,5}S^e$ 
and are in reasonable agreement with the stabilization calculations ~\cite{Hara1989} for $^{3}S^e$, $^{3}P^o$, and $^{1,5}P^o$.
\add{The non-radiative decay widths are small and
the complex coordinate rotation trajectories of the resonance states 
are located close to the real axis. While the trajectories do not show a clear pole except for $v=0^\ast$ state in $^{1,5}P^o$,
we estimate the resonance widths by the 3 times of the standard deviation of the complex energies near the real axis for several $\theta$. 
The estimated resonance widths are less than $\lesssim$10$^{-7}$ hartree or 10 $\murm$eV, which is 
smaller than the significant digits of resonance energies listed in Table~\ref{tbl_ddmueng_comp}.
In the subsequent calculations of the radiative decay, we verified that the x-ray spectra and radiative decay widths are
reproduced even if we use resonance state wavefunctions corresponding to the slightly different eigenenergies within the non-radiative widths.
}

\begin{figure}[tb]
\begin{center}
\resizebox{0.5\textwidth}{!}{%
\includegraphics[bb=0 0 368 453]{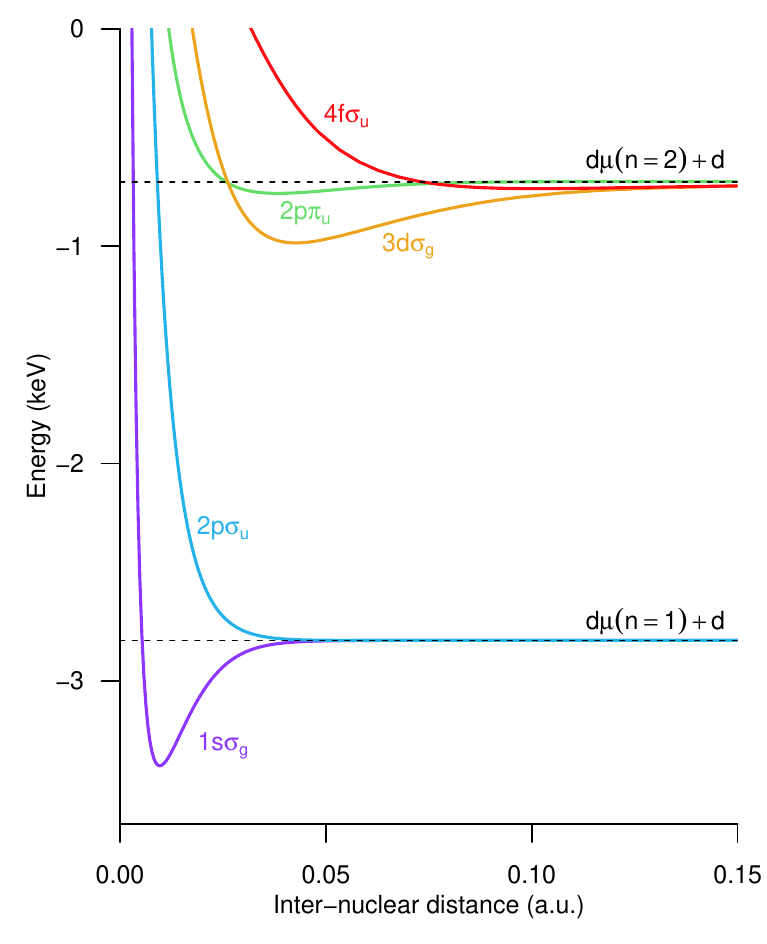}}%
\caption{
Adiabatic potential energy curves for dd$\murm$. The lower and upper dashed lines correspond to d$\murm(n=1)$ + d and d$\murm(n=2)$ + d, respectively.
}
\label{f_BO}
\end{center}
\end{figure}

By applying the Born-Oppenheimer approximation to the d+d+$\murm$ system, three adiabatic potential energy curves, $3d\sigma_g$, $4f\sigma_u$, and $2p\pi_u$, are obtained.
They are attractive at long inter-nuclear distances and approach asymptotically to the d$\murm(n=2)$ + d threshold energy.
Both $3d\sigma_g$ and $4f\sigma_u$ potential energy curves are inversely proportional to the square of the distance at long distances.
In contrast, the $2p\pi_u$ potential is inversely proportional to the fourth power of the distance. 
Figure~\ref{f_BO} illustrates these adiabatic potential energy curves as well as $1s\sigma_g$ and $2p\sigma_u$ curves.
The resonance states of $^{1,5}S^e$ and $^{3}P^o$ belong to the $3d\sigma_g$ adiabatic potential energy curve.
The other resonance states of $^{3}S^e$ and $^{1,5}P^o$, except for the $v_\mathrm{r}=0^\ast$ state of $^{1,5}P^o$, belong to 
the $4f\sigma_u$ adiabatic potential energy curve. The $v_\mathrm{r}=0^\ast$ state of $^{1,5}P^o$ is referred to as an even-parity bound state~\cite{Hara1989}
and is supported by the $2p\pi_u$ potential. 

The degeneracy of d$\murm(2s)$ and d$\murm(2p)$ causes Stark mixing owing to the charge of the other d, resulting in an infinite series
of the resonance energy levels, called the dipole series. The resonance states of the dipole series are 
supported by an attractive ion-dipole interaction, 
which is proportional to the inverse of the square of the d$\murm$--d distance.
However, as the degeneracy of $n=2$ levels of d$\murm$ is resolved mainly by vacuum polarization, 
the resonance series is truncated at high vibration levels. 
As vacuum polarization results in an approximately 0.2 eV $2s$--$2p$ interval, we enlarged the expansion range of the basis functions to $\sim 240$ m.a.u.
such that a shallow resonance state having the binding energy of sub-electronvolts can be described with reasonable accuracy.

\begin{figure}[tb]
\begin{center}
\resizebox{0.5\textwidth}{!}{%
\includegraphics[bb=0 0 368 283]{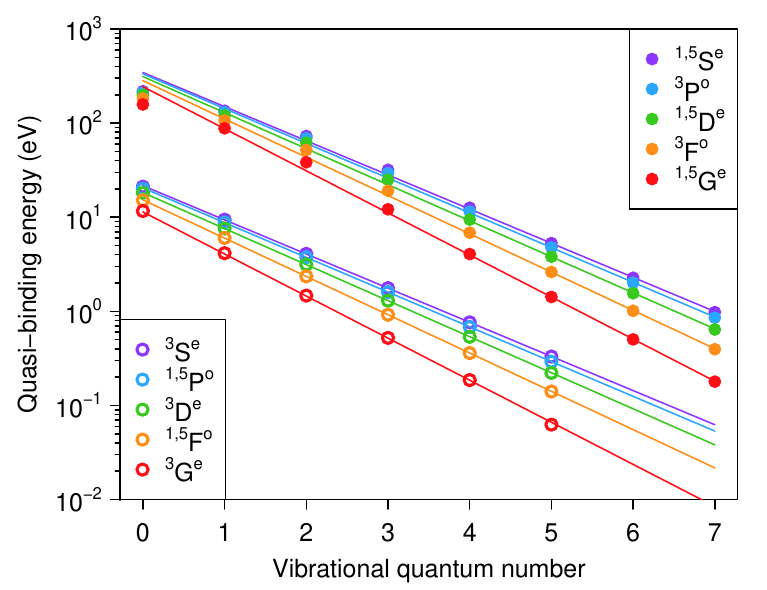}}%
\caption{
Quasi binding energies of $\ddmr$ resonance states are presented against the corresponding vibrational quantum numbers. 
The lines are fitted using an exponential function.
}
\label{f_dipole_ser}
\end{center}
\end{figure}

The energy intervals of the dipole resonance states near the threshold energy obey the following simple analytical formula.
\begin{align}
\frac{\varepsilon_{J,v_\mathrm{r}}}{\varepsilon_{J,v_\mathrm{r}+1}} = \exp\left({\frac{2\pi}{\alpha_{J}}}\right),
\end{align}
where $\alpha_J$ is a constant that depends on the total angular momentum $J$ as
\begin{align}
\label{alpha_J}
\alpha_J = \sqrt{ -\frac{1}{4} - J^2+J+1 + (2J+1) \gamma_J }.
\end{align}
Here,
\begin{align}
\gamma_J = \sqrt{\left(\frac{6\mu_{\mathrm{d}\murm,\mathrm{d}}}{\mu_{\mathrm{d},\murm}(2J+1)}\right)^2+1},
\end{align}
where $\mu_{\mathrm{d},\murm}$ and $\mu_{\mathrm{d}\murm,\mathrm{d}}$ are reduced masses of d + $\murm$ and d$\murm$ + d, respectively. 
The values of $\alpha_J$ for the dd$\murm^\ast$ resonances are
$\alpha_{J=0} = 1.197$, $\alpha_{J=1} = 1.176$, $\alpha_{J=2} = 1.134$, $\alpha_{J=3} = 1.067$, and $\alpha_{J=4} = 0.971$.

Figure~\ref{f_dipole_ser} presents the quasi-binding energies $\varepsilon_{J,v_\mathrm{r}}$ versus $v_\mathrm{r}$.
The dashed lines are fitted by
\begin{align}
\varepsilon_{J,v_\mathrm{r}} = A \exp\left(-{\frac{2\pi}{\alpha_{J}}}v_\mathrm{r}\right).
\end{align}
Using the analytical values of $\alpha_J$ in Eq.~(\ref{alpha_J}), we optimized $A$ for the high vibrational states as follows:
$4\leq v_\mathrm{r}\leq7$ for $^{1,5}S^e$, $^3P^o$, $^{1,5}D^e$, $^3F^o$, and $^{1,5}G^e$ resonances
and $3\leq v_\mathrm{r}\leq5$ for $^{3}S^e$, $^{1,5}P^o$, $^{3}D^e$, $^{1,5}F^o$, and $^{3}G^e$ resonances.
The energy interval of these high vibrational states demonstrates good agreement with the analytical expression,
that is, the obtained dd$\murm^\ast$ resonance energies are considered to be accurate.

\subsection{Convergence of x-ray spectrum}

\begin{figure}[tb]
\begin{center}
\resizebox{0.45\textwidth}{!}{%
\includegraphics[bb=0 0 311 283]{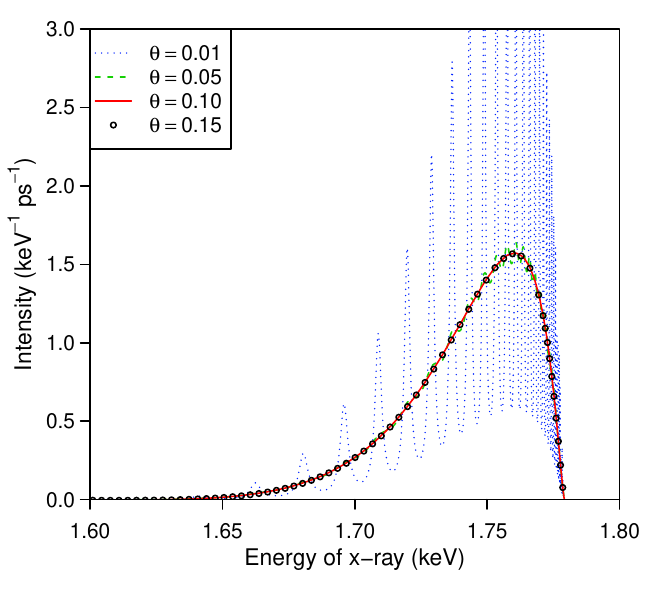}}%
\caption{
Convergence of x-ray spectrum for the $^{1,5}S^e\ v_\mathrm{r}=0$ state 
against the complex coordinate rotation angle $\theta$.
}
\label{f_raddis_conv1}
\end{center}
\end{figure}

We examine the convergence of the x-ray spectra from the decay of the resonance states into a continuum using Eq. ~(\ref{e_raddec}).
Figure~\ref{f_raddis_conv1} presents the convergence of the x-ray spectrum from the $^{1,5}S^e, v_\mathrm{r}=0$ state
against the complex coordinate rotation angle $\theta$. 
At $\theta=0.01$, the spectrum exhibits an oscillation against the x-ray energy, while
the oscillation disappears as $\theta$ increases. 
The antinodes of the oscillation appear when the denominator $(E_{J_\mathrm{f},v}(\theta)-E_\mathrm{f})$ in Eq. (\ref{e_raddec})
is approximately zero at small $\theta$
while the
$E_{J_\mathrm{f},v}(\theta)$ is a complex value with a non-zero imaginary part such that $(E_{J_\mathrm{f},v}(\theta)-E_\mathrm{f})^{-1}$ is non-singular 
at all $E_\mathrm{f}$ values in the continuum state.
The present calculation confirmed that $\theta\geq 0.1$ results in a well-converged x-ray spectrum.

\begin{figure}[tb]
\begin{center}
\resizebox{0.45\textwidth}{!}{%
\includegraphics[bb=0 0 311 283]{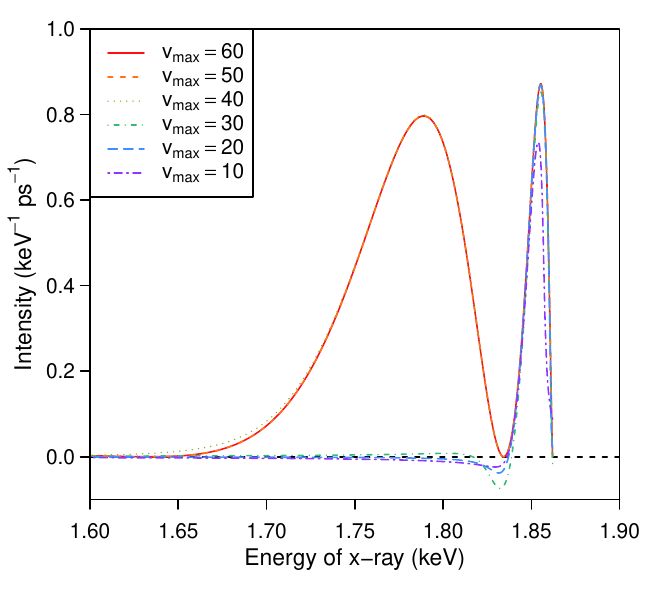}}%
\caption{
Convergence of x-ray spectrum for $^{1,5}S^e\ v_\mathrm{r}=1$ state 
against the maximum number of eigen functions $v_\mathrm{max}$.
}
\label{f_raddis_conv2}
\end{center}
\end{figure}

In terms of the accuracy of the wave functions of the resonance states, 
we use a sufficient number of basis functions and examine the Gaussian size parameters in Eq.~(\ref{wf1}) 
such that the well converged energies are obtained for low $v_\mathrm{r}$ states and physically meaningful dipole series are reproduced for high $v_\mathrm{r}$ states. 

To examine the validity of the approximation for the calculation of the x-ray spectra described by Eq.~(\ref{e_raddec}),
we examine the convergence of the x-ray spectra by increasing the number of eigen functions $\{{\psi}_{J_\mathrm{f},v}(\theta)\}$, $v_\mathrm{max}$, used in Eq. ~(\ref{e_raddec}).
We sorted $\{\psi_{J_\mathrm{f},v}(\theta)\}$ in ascending order of $\mathrm{Re}\,E_{J_\mathrm{f},v}(\theta)$ and increased $v_\mathrm{max}$.
Figure~\ref{f_raddis_conv2} presents an example of the convergence behavior against $v_\mathrm{max}$ for the $^{1,5}S^e\ v_\mathrm{r}=1$ state.
In this calculation, $\theta=0.15$ is used. 
We conclude that spectra are converged against the $v_\mathrm{max}\geq50$.
A drastic change occurs between $v_\mathrm{max}=30$ and $40$ where a low energy peak appears.
Real parts of the complex eigen energies, i.e. kinetic energy $K_\mathrm{rel}\cos2\theta$ in Eq~(\ref{rel_kinetic_rot}),
$\mathrm{Re}\, E_{J_\mathrm{f},v=30}(\theta) - E_{\mathrm{th}}^{(n=1)}=0.000\,013$ m.a.u. and $\mathrm{Re}\, E_{J_\mathrm{f},v=40}(\theta)- E_{\mathrm{th}}^{(n=1)}=0.027\,767$ m.a.u., 
are associated with the x-ray energies $E_\gamma$ of $1.86$ keV and $1.71$ keV, respectively, with $E_\gamma = E_{J,v}^\mathrm{(R)} - \mathrm{Re}\, E_{J_\mathrm{f},v}(\theta)$.
The presented convergence behavior suggests that 
$\{\psi_{J_\mathrm{f},v}\}$ for $v\leq 30$ are not enough for the closure relation in Eq.~(\ref{closure});
as the members of $\{\psi_{J_\mathrm{f},v}\}$ increase, the energy range of the x-ray spectra becomes wider.

\begin{figure}[tb]
\begin{center}
\resizebox{0.45\textwidth}{!}{%
\includegraphics[bb=0 0 311 283]{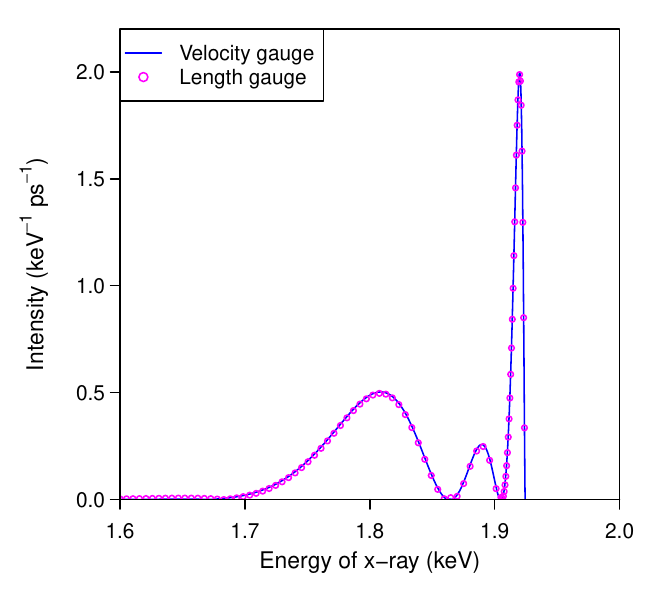}}%
\caption{
Comparison of the velocity and length gauge calculations for the $^{1,5}S^e\ v_\mathrm{r}=2$ state.
}
\label{f_raddis_conv3}
\end{center}
\end{figure}

We also examine the x-ray spectrum by comparing the length- and velocity-gauge calculations.
Figure~\ref{f_raddis_conv3} presents an example of the length- and velocity-gauge calculations for the $^{1,5}S^e\ v_\mathrm{r}=2$ state using the same basis functions and 
complex rotation angle.
The x-ray spectra obtained using two kinds of gauges exhibit excellent agreement with each other, confirming the validity of our calculations.

\subsection{Characteristic shape of x-ray spectrum}

\begin{figure*}[t]
\begin{center}
\resizebox{\textwidth}{!}{%
\includegraphics[bb=0 0 850 850]{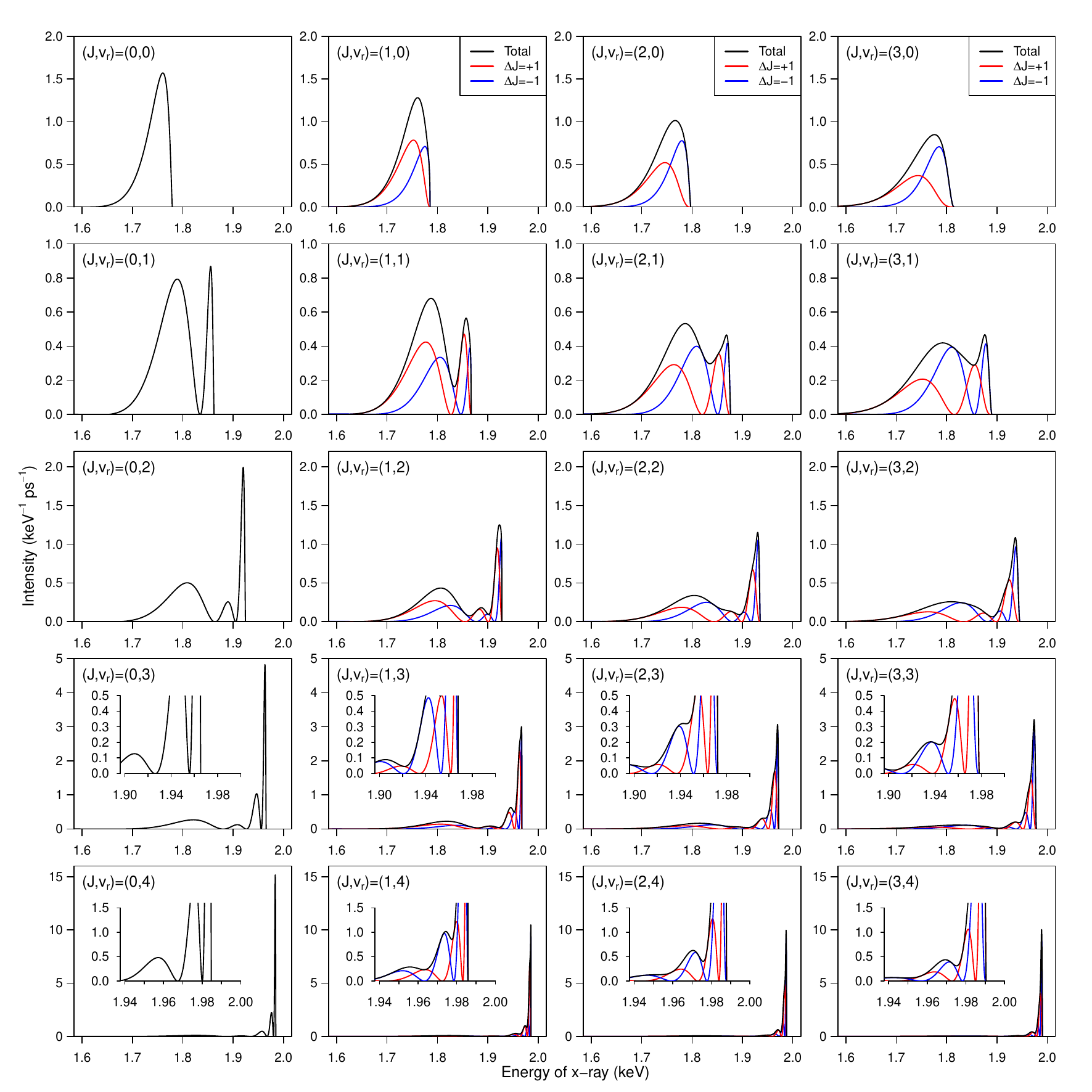}}%
\caption{
X-ray spectra of radiative decay into continuum states from resonance states belonging to
$^{1,5}S^e$, $^{  3}P^o$, $^{1,5}D^e$, and $^{  3}F^o$ series. 
}
\label{f_spec_serA1}
\end{center}
\end{figure*}

\begin{figure*}[t]
\begin{center}
\resizebox{\textwidth}{!}{%
\includegraphics[bb=0 0 850 481]{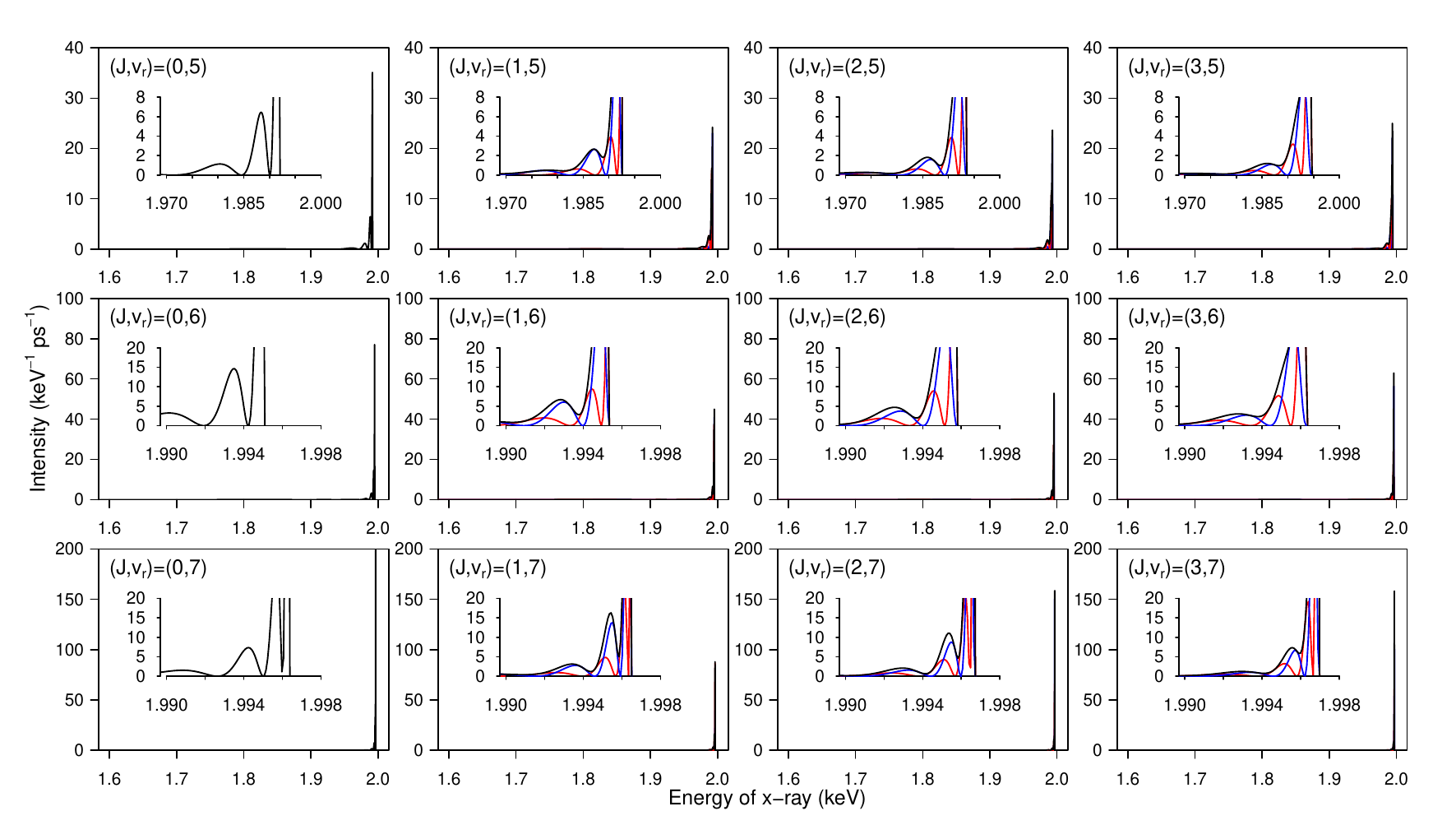}}%
\caption{
\textit{(Continued from Fig.~\ref{f_spec_serA1}.)}
}
\label{f_spec_serA2}
\end{center}
\end{figure*}

\begin{figure*}[t]
\begin{center}
\resizebox{\textwidth}{!}{%
\includegraphics[bb=0 0 850 850]{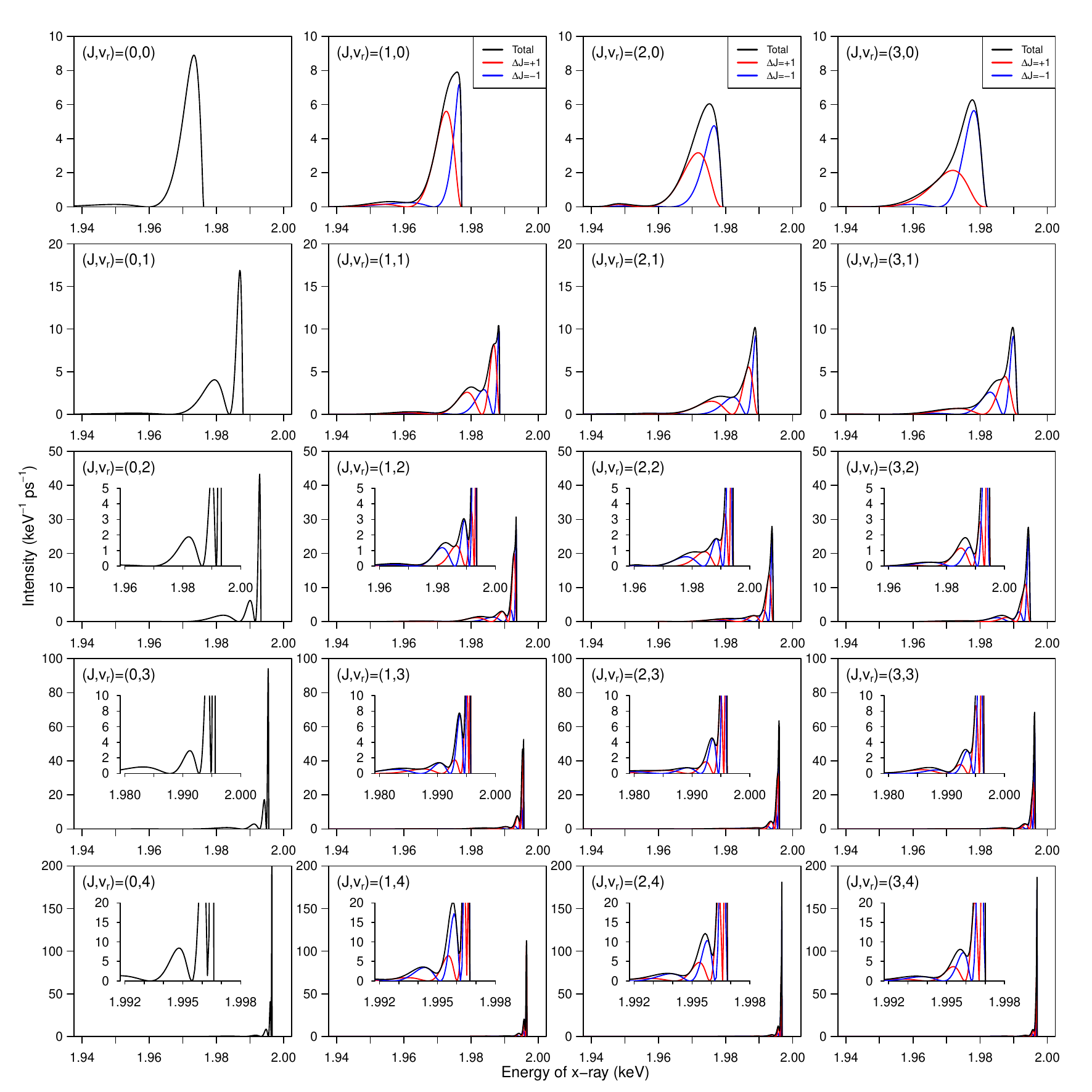}}%
\caption{
X-ray spectra of radiative decay into continuum states from resonance states belonging to
$^{3}S^o$, $^{1,5}P^e$, $^{3}D^o$, and $^{1,5}F^e$ series. 
}
\label{f_spec_serB}
\end{center}
\end{figure*}

Figures~\ref{f_spec_serA1}, \ref{f_spec_serA2}, and \ref{f_spec_serB} present a summary of the x-ray spectra of radiative decay to the continuum states
for each resonance state.
The maximum x-ray energy corresponds to the energy interval between the resonance energy $E_{J,v_\mathrm{r}}^{\mathrm{(R)}}$ 
and d$\murm(n=1)$ + d dissociation threshold energy $E_\mathrm{th}^{(n=1)}$ as follows:
\begin{align}
E_\gamma \leq E_{J,v_\mathrm{r}}^{\mathrm{(R)}} - E_\mathrm{th}^{(n=1)}.
\end{align}
For the $J=0$ resonance states, radiative decay results in a $J_\mathrm{f}=1$ continuum state
within the dipole approximation. As we do not include spin-dependent interactions between the two deuterons,
the total nuclear spin angular momentum $S_\mathrm{nucl}$ is conserved throughout the radiative decay.
For $J\geq 1$ resonance states, the radiative decay results in $J_\mathrm{f}=J\pm1$ continuum states.
We distinguish between the $J$-increasing decay and $J$-decreasing decay by denoting $\Delta J=\pm1$.

As shown in Fig.~\ref{f_spec_serA1}, which summarizes the spectra of the resonance states belonging to $3d\sigma_g$ series,
the x-ray spectrum from $^{1,5}S^e$ resonance states increases the number of nodes with an increase in $v_\mathrm{r}$.
A similar trend can be observed for higher angular momenta, whereas the node structure is evident only for the partial spectra
and disappears from the total x-ray spectrum because of the shift in the node positions. 
The decrease of x-ray spectra of $\Delta J=+1$ decay near the maximum x-ray energy is steeper than those of $\Delta J=-1$ 
because of centrifugal repulsion between the decay fragments.

Figure~\ref{f_spec_serA2} summarizes the x-ray spectra of the $v_\mathrm{r}\geq5$ resonance states of the $3d\sigma_g$ series. 
Again, the number of the nodes in the x-ray spectrum increases as an increase of $v_\mathrm{r}$.
The intensity near the maximum x-ray energy increases as an increase of $v_\mathrm{r}$.

For the resonance states of the $4f\sigma_u$ series, as shown in Fig. ~\ref{f_spec_serB}, a similar trend can be observed, 
i.e., the number of nodes increases with $v_\mathrm{r}$.
It should be noted that the $4f\sigma_u$ resonances have a shallower quasi-binding energy than the $3d\sigma_g$ resonances
with the same $v_\mathrm{r}$, which results in higher maximum x-ray energy than the $3d\sigma_g$ resonances.

To investigate the characteristic shape of the x-ray spectrum, we consider
the vertical transition between the resonance and continuum states.
Radial distribution functions of the resonance states are calculated by 
\begin{align}
\label{rdf}
D(R'_1) = {R'}^2_1 \left\langle \Psi_\mathrm{R} \bigg| \frac{\delta(R_1-R'_1)}{4\pi R_1^2} \bigg| \Psi_\mathrm{R} \right\rangle,
\end{align}
where $R_1$ denotes the distance between the center-of-mass of d$\murm$ and d, as shown in Fig.~\ref{f_coord}.
Here, we use the unrotated wave function $\Psi_\mathrm{R}$. 
This approximation is appropriate when the energy widths of the resonance states are much smaller than the resonance energy.
For the resonance states of $\ddmr$, the resonance width against the non-radiative decay is 
sufficiently small (less than $10$ $\murm$eV~\cite{Kilic2004}). 
We, therefore, can treat $\Psi_\mathrm{R}(\theta=0)$ as a resonance-state wave functions in Eq.~(\ref{rdf})
to discuss the origin of the characteristic shapes of the x-ray spectra. 
 
\begin{figure*}[t]
\begin{center}
\resizebox{0.9\textwidth}{!}{%
\includegraphics[bb=0 0 566 368]{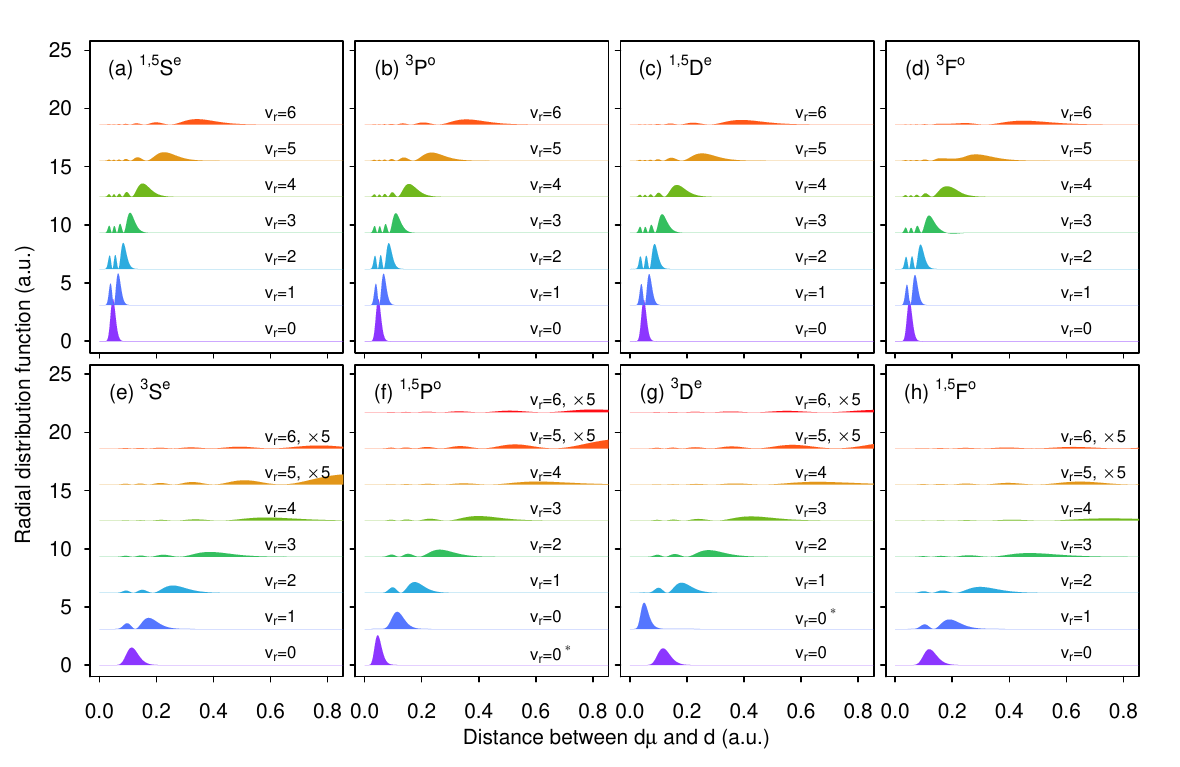}}%
\caption{
Radial distribution functions of $\ddmr$ are presented as a function of the distance between the center-of-mass of d$\murm$ and d.
}
\label{f_corr}
\end{center}
\end{figure*}

Figure~\ref{f_corr} presents the radial distribution functions of the resonance states.
As expected, the radial distribution functions increase their number of nodes as the vibrational quantum number $v_\mathrm{r}$ increases.
The resonance states in the $3d\sigma_g$ series ($^{1,5}S^e$, $^3P^o$, $^{1,5}D^e$, $^3F^o$, and $^{1,5}G^e$) exhibit a relatively narrow distribution 
in comparison with those in the $4f\sigma_u$ series ($^{3}S^e$, $^{1,5}P^o$, $^{3}D^e$, $^{1,5}F^o$, and $^{3}G^e$).
The difference among the radial distributions is consistent with the adiabatic potential energy curves presented in Fig.~\ref{f_BO}, 
i.e., the $4f\sigma_u$ adiabatic potential energy curve is more repulsive at short distances and has a shallower well
compared to the $3d\sigma_g$ adiabatic potential energy curve.
The resonance states denoted by $v_\mathrm{r}=0^\ast$ in $^{1,5}P^o$ and $^{3}D^e$ have significantly narrow distributions. 
These resonance states belong to the $2p\pi_u$ series, and their narrow distribution is consistent with the short-range adiabatic potential energy curve of $2p\pi_u$.

According to Ref.~\cite{Lindroth2003}, the x-ray spectrum of the radiative decay into the continuum can be qualitatively understood 
as a vertical transition from the resonance energy level
to the dissociative $2p\sigma_u$ potential energy curve.
The low-energy boundary of the x-ray spectrum can be roughly estimated by the repulsive $2p\sigma_u$ potential energy curve $V_{2p\sigma_u}(R_1)$ as follows: 
\begin{align}
E_\gamma \geq E_{J,v_\mathrm{r}}^{\mathrm{(R)}} - V_{2p\sigma_u}(R_1^\mathrm{IT}),
\end{align}
where $R_1^\mathrm{IT}$ denotes the inner turning distance of the resonance state. 
As the adiabatic potential $V_{2p\sigma_u}(R_1)$ is repulsive on the foot of the potential and asymptotically converges to the threshold energy of d$\murm(1s)$ + d,
the narrow radial distribution of the resonance state tends to result in a widely distributed x-ray spectrum. 
The diffused structure, in turn, tends to produce a sharp x-ray spectrum near the maximum x-ray energy, which is consistent with the calculated spectra 
presented in Figs.~\ref{f_spec_serA1}, \ref{f_spec_serA2}, and \ref{f_spec_serB}.

\begin{figure}[tb]
\begin{center}
\resizebox{0.5\textwidth}{!}{%
\includegraphics[bb=0 0 368 481]{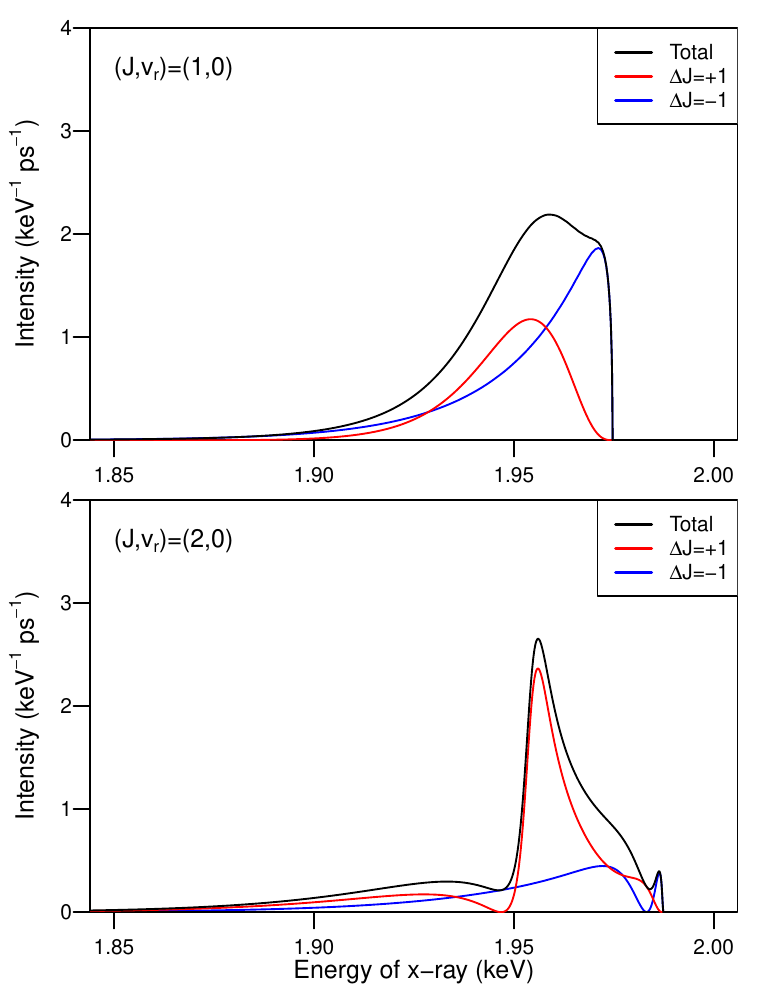}}%
\caption{
X-ray spectra from $v_\mathrm{r} = 0^\ast$ states in $^{1,5}P^o$ (upper panel) and $v_\mathrm{r} = 0^\ast$ (lower panel).
}
\label{f_spec_serC}
\end{center}
\end{figure}

Figure~\ref{f_spec_serC} presents the x-ray spectra of the two states of $v_\mathrm{r}=0^\ast$ supported by the $2p\pi_u$ adiabatic potential energy curve. 
As listed in Table~\ref{tbl_ddmueng_comp}, the quasi-binding energy of $v_\mathrm{r} = 0^\ast$ in $^{1,5}P^o$ is 22.6 eV which differs only by 2.5 eV from the
$v_\mathrm{r}= 0$ state in $^{1,5}P^o$ ($4f\sigma_u$ series). However, on comparing the x-ray spectrum of Fig.~\ref{f_spec_serC} $(J,v_\mathrm{r})=(1,0)$ 
with Fig.~\ref{f_spec_serB} $(J,v_\mathrm{r})=(1,0)$, it is observed that the x-ray spectrum from $v_\mathrm{r}= 0^\ast$ in the $2p\pi_u$ series is broader 
than the $v_\mathrm{r}=0$ state in the $4f\sigma_u$ series. 
A similar trend is observed for the $v_\mathrm{r} = 0^\ast$ state in $^3D^e$. 

\subsection{Kinetic energy distribution of the decay fragments}

\begin{figure}[tb]
\begin{center}
\resizebox{0.5\textwidth}{!}{%
\includegraphics[bb=0 0 396 396]{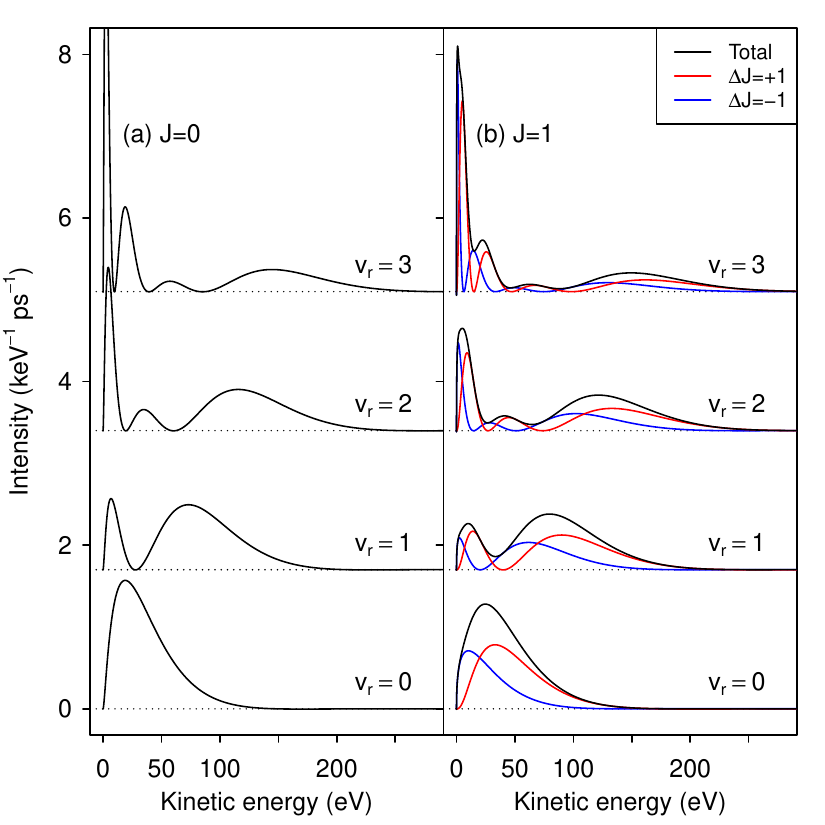}}%
\caption{
Kinetic energy distribution of the decay fragments in the center-of-mass framework for $0\leq v_\mathrm{r} \leq3$ resonance states of the $3d\sigma_g$ series.
}
\label{f_kinetic_dist}
\end{center}
\end{figure}

The kinetic energy distribution of decay fragments is an interesting subject in the context of the muon catalyzed fusion
because it is necessary to evaluate epi-thermal muonic atoms~\cite{Cohen1986,Markushin1993,Fujiwara2000}.
After the radiative decay, a part of the released energy $E_{J,v_\mathrm{r}}^{\mathrm{(R)}}-E_\mathrm{th}^{(n=1)}$ becomes the kinetic energy of the decay fragments.
The $E_{J,v_\mathrm{r}}^{\mathrm{(R)}}-E_\mathrm{th}^{(n=1)}$ is expressed as a sum of
the x-ray energy ($E_\gamma$) and kinetic energy ($K$) of the d$\murm(1s)$--d relative motion:
\begin{align}
E_{J,v_\mathrm{r}}^{\mathrm{(R)}}-E_\mathrm{th}^{(n=1)} = E_\gamma + K.
\label{divide}
\end{align}
As observed in Figs.~\ref{f_spec_serA1}, \ref{f_spec_serA2}, and \ref{f_spec_serB},
each partial x-ray spectrum corresponding to the $\Delta J=\pm1$ decay has nodes.
The presence of the nodes in x-ray spectra corresponds to the nodes of the kinetic energy distribution of the relative motion of the decay fragments.
Figure~\ref{f_kinetic_dist} shows the kinetic energy distribution from $0\leq v_\mathrm{r} \leq 3$ resonance states 
of $J=0$ and $1$ supported by the $3d\sigma_g$ potential curve.
The kinetic energy distribution shown in Fig.~\ref{f_kinetic_dist} indicates that the decay fragments have several tens eV 
even though most of the total released energy $E_{J,v_\mathrm{r}}^{\mathrm{(R)}}-E_\mathrm{th}^{(n=1)}$ are carried out by the x-ray.
For further investigation, we denote the nodes of kinetic energy distribution by $K_\mathrm{node}$. 
The $K_\mathrm{node}$ depends on the angular momentum of the relative motion of decay fragments $L_\mathrm{f}=J_\mathrm{f}$.

\begin{figure}[tb]
\begin{center}
\resizebox{0.40\textwidth}{!}{%
\includegraphics[bb=0 0 208 299]{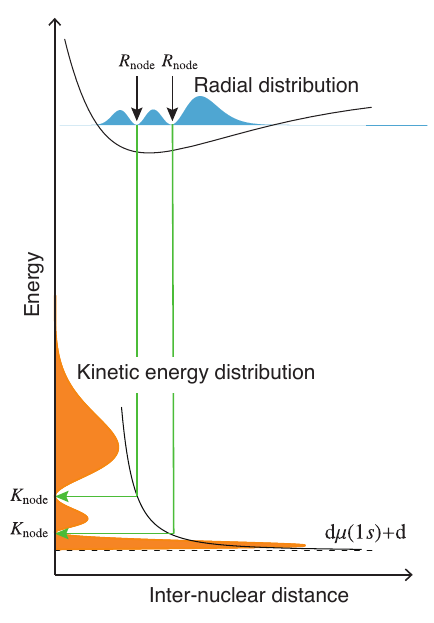}}%
\caption{
Schematic of the relationship between the node of the kinetic energy distribution $K_\mathrm{node}$
and the node of radial distribution $R_\mathrm{node}$ of the radial distribution function of the resonance states.
Assuming vertical transitions onto the repulsive potential energy curve of the decay fragments, 
the shorter $R_\mathrm{node}$ can correspond to a higher $K_\mathrm{node}$.
}
\label{f_nodes_expl}
\end{center}
\end{figure}

As shown in Fig.~\ref{f_corr}, the radial distribution functions in Eq.~(\ref{rdf}) have nodes at certain distances $R_1=R_\mathrm{node}$. 
Assuming vertical transitions from the resonance state to a lower dissociative potential energy curve,
the kinetic energy node $K_\mathrm{node}$ can be associated with the distance node $R_\mathrm{node}$ of the radial distribution function,
as illustrated in Fig.~\ref{f_nodes_expl}.

\begin{figure}[tb]
\begin{center}
\resizebox{0.5\textwidth}{!}{%
\includegraphics[bb=0 0 368 453]{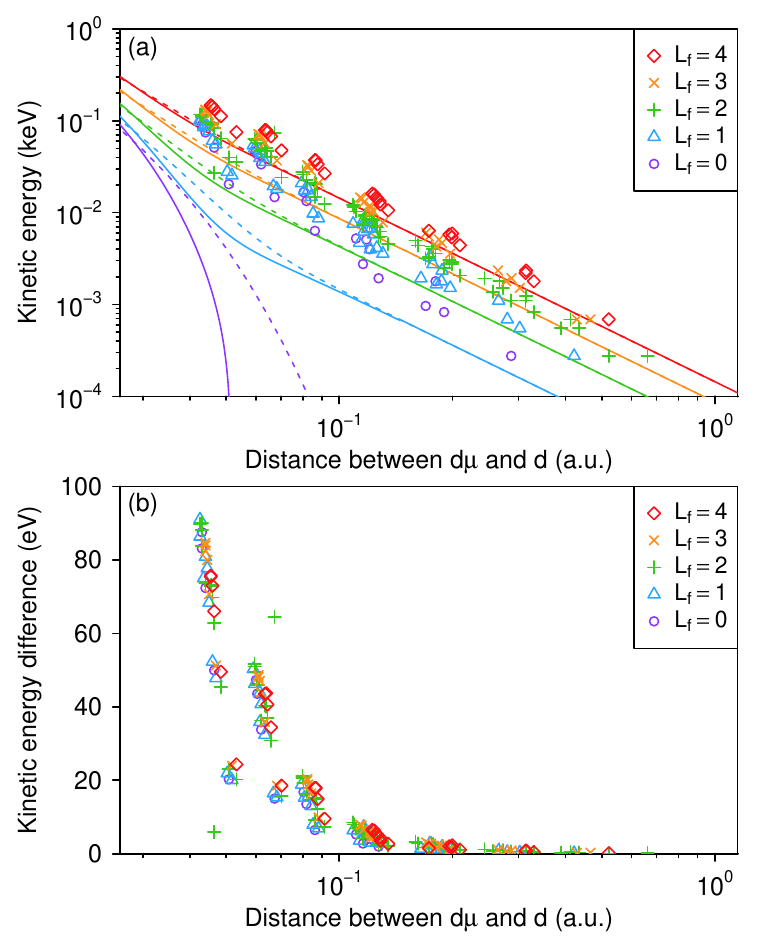}}%
\caption{
(a) Nodes of kinetic energy distribution $K_\mathrm{node}$ of relative motion of the decay fragments 
are plotted by points against the node of radial distribution $R_\mathrm{node}$ of the resonance states.
The solid lines are given by $K_\mathrm{node}^\mathrm{(ad)}(R_\mathrm{node})$, and the dashed line is given by $K_\mathrm{node}^\mathrm{(st)}(R_\mathrm{node})$.
(b) The differences between $K_\mathrm{node}$ and $K_\mathrm{node}^\mathrm{(ad)}(R_\mathrm{node})$ are shown against $R_\mathrm{node}$.
}
\label{f_nodes}
\end{center}
\end{figure}

We analyze the relationship between $K_\mathrm{node}$ and $R_\mathrm{node}$ as shown in  
Fig.~\ref{f_nodes}(a) by specifying $L_\mathrm{f}$.  
It is apparent that, at the same $R_\mathrm{node}$, $K_\mathrm{node}$ is higher at a higher $L_\mathrm{f}$.
To quantitatively investigate the $K_\mathrm{node}$--$R_\mathrm{node}$ relationship, we introduce an adiabatic estimation of $K_\mathrm{node}$ using
\begin{align}
K_\mathrm{node}^\mathrm{(ad)}(R_\mathrm{node}) = V_{2p\sigma_u}(R_\mathrm{node})
                                               + \frac{L_\mathrm{f}(L_\mathrm{f}+1)}{2 \mu_{d\murm,d} R_\mathrm{node}^2}, 
\label{eq_kinetic_ad}
\end{align} 
where $V_{2p\sigma_u}$ is the $2p\sigma_u$ adiabatic potential energy curve measured from the threshold energy of d$\murm(1s)+$d,
and $\mu_{d\murm,d}$ is the reduced mass of d$\murm$ and d.
Equation~(\ref{eq_kinetic_ad}) indicates that a higher $L_\mathrm{f}$ results in a higher centrifugal barrier between the decay fragments
and a higher $K_\mathrm{node}$ at the same distance node.

The estimated $K_\mathrm{node}^\mathrm{(ad)}$ values are presented in Fig.~\ref{f_nodes}(a) by solid lines.
In the case $L_\mathrm{f}=0$, the estimated $K_\mathrm{node}^\mathrm{(ad)}(R_\mathrm{node})$ fails to explain the $K_\mathrm{node}$ derived from the 
three-body calculation. 
In particular, $V_{2p\sigma_u}$ has a shallow attractive well at a large distance; 
therefore, $K_\mathrm{node}^\mathrm{(ad)}(R_\mathrm{node})$ becomes negative at large distances.
In the non-zero $L_\mathrm{f}$ case, the decreasing behavior of $K_\mathrm{node}$ against $R_\mathrm{node}$ 
demonstrates a trend similar to the $E_\mathrm{node}^\mathrm{(ad)}(R_\mathrm{node})$; however, 
even at $L_\mathrm{f}=4$, $K_\mathrm{node}^\mathrm{(ad)}(R_\mathrm{node})$ does not coincide with the $K_\mathrm{node}$. 

We can assume a simpler model of the kinetic energy node as
\begin{align}
K_\mathrm{node}^\mathrm{(st)}(R_\mathrm{node}) 
  = \frac{\exp(-2R_\mathrm{node}/a_\mu )}{R_\mathrm{node}} + \frac{L_\mathrm{f}(L_\mathrm{f}+1)}{2 \mu_{d\murm,d} R_\mathrm{node}^2}, 
\label{eq_kinetic_st}
\end{align} 
where the first term denotes the screened Coulomb potential between d$\murm(1s)$ and d. $a_\mu$ denotes a Bohr radius of d$\murm(1s)$.
In contrast with $V_{2p\sigma_u}(R_\mathrm{node})$, the screened Coulomb potential is repulsive to any $R_\mathrm{node}$;
however,  $K_\mathrm{node}^\mathrm{(st)}(R_\mathrm{node})$ fails to reproduce $K_\mathrm{node}$ in the three-body calculation,
as indicated by the dashed lines in Fig.~\ref{f_nodes}(a).
The discrepancy between the $K_\mathrm{node}$ and $K_\mathrm{node}^\mathrm{(ad)}$ is highlighted in Fig.~\ref{f_nodes}(b), which
indicates that the difference reaches nearly 90 eV at a short distance for any $L_\mathrm{f}$. 

Therefore, the radiative decay into the continuum does not occur \textit{on to} the 
adiabatic potential energy curve.
The decay fragments have much higher kinetic energies than the adiabatic energy curve provides.
Within the authors' knowledge, the muonic molecules in the resonance states has not been considered as a heating source of muonic atoms in the \mcf\ models.
We stress that the radiative decay of $\ddmr$ can play an important role in producing epi-thermal muonic atoms and
the non-adiabatic effects are indispensable for determining the kinetic energy distribution (and x-ray spectra).

When the dd$\murm^\ast$ resonance state forms via the Vesman mechanism as shown in Eq.~(\ref{vesman2}),
dd$\murm^\ast$ is in the electron cloud of the host molecule $[(\ddmr)dee]$. 
The rates of the Auger transition,
\begin{align}
[(\ddmr)\mathrm{dee}] \to \ddmr + \mathrm{e} + \mathrm{D},
\end{align}
are much higher than the radiative decay rates~\cite{Froelich1995}, the dd$\murm^\ast$ first emits an Auger electron and 
is deexcited to lower rovibrational states (Auger decay).
Because the Auger decay requires an energy interval of 15.4 eV to overcome the ionization energy of hydrogen molecules,
the radiative decay dominates for $0\leq v_\mathrm{r} \leq3$
resonance states of the $3d\sigma_g$ series.
A more detailed investigation of the Auger processes will be presented in our forthcoming publication.

\subsection{Radiative decay into bound state}

The total rates of radiative decay into the continuum state can be obtained by integrating the x-ray spectrum 
${\mathrm{d}\varGamma_\gamma}/{\mathrm{d}E_\gamma}$ over $E_\gamma$. 
In addition to the decay into the continuum state, we investigate the radiative decay into the bound states (RB decay),
\begin{align}
\ddmr(J,v_\mathrm{r}) \to \ddm(J_\mathrm{f},v_\mathrm{b}) + \gamma,
\end{align}
and into the other resonance states (RR' decay),
\begin{align}
\ddmr(J,v_\mathrm{r}) \to \ddmr(J_\mathrm{f},v'_\mathrm{r}) + \gamma.
\end{align}
These rates can be calculated as
\begin{equation}
\varGamma_\mathrm{RB(RR')} = \frac{4}{3}\alpha^3 E_\gamma^3\,
 \langle \bar{\Psi}_\mathrm{B(R')}(\theta)  | \mathbf{d}(\theta) |  \Psi_\mathrm{R}(\theta) \rangle^2,
\label{rate_rrrb}
\end{equation}
where 
$\Psi_\mathrm{R}(\theta)$ is a complex rotated wave function of the resonance states,
$\Psi_\mathrm{B}(\theta)$ is that of the final bound state, and
$\Psi_\mathrm{R'}(\theta)$ is that of the other resonance state.
$E_\gamma$ corresponds to the energy interval between the initial and final state energies.
\add{The $\varGamma_\mathrm{RB(RR')}$ is in principle complex value. 
The real part can be interpreted as the observable of the expectation value and the imaginary part is considered to be 
its uncertainty. In the present calculation, owing to the small non-radiative decay widths of the resonance states, 
the real part of the $\varGamma_\mathrm{RB(RR')}$ is much larger than the imaginary part 
(typically $|\mathrm{Im}\,\varGamma_\mathrm{RB(RR')}/\mathrm{Re}\,\varGamma_\mathrm{RB(RR')}|<10^{-3}$).
Thus, in the following discussion, we only report the real part of the $\varGamma_\mathrm{RB(RR')}$.
}

Table~\ref{tbl_rr-rb} lists $\varGamma_\mathrm{RB(RR')}$ in the descending order. 
These rates agree well with each other in both the length and velocity gauges.
The largest rates are found for RB decay from the $v_\mathrm{r}=0^\ast$ state in $^{3}D^{e}$ and $^{1,5}P^{o}$.
The $2p\pi_u$ adiabatic potential curve supports these resonance states
and their radial distribution function has a peak at a relatively short distance, namely at $0.045$ bohr for $v_\mathrm{r}=0^\ast$ state of $^{1,5}P^{o}$ 
and $0.050$ bohr for that of $^{3}D^{e}$, as shown in Fig.~\ref{f_corr}(f) and (g).
The $v_\mathrm{r}=0$ states in $^{3}S^e$ and $^{3}D^e$ also exhibit relatively larger rates than the others
while the radial distribution functions of these resonance states have peaks longer than $0.11$ bohr.
Both the states result in the bound state $(J,v_\mathrm{b})=(1,1)$, which is a loosely bound state 
having a binding energy of only 1.97 eV and a diffused radial distribution function.
Thus, the concentrated inter-nuclear wave functions in resonance states or the diffused wave functions in the bound states 
lead to exhibit relatively high RB decay rates.

\begin{table}[t]                
\begin{ruledtabular}                
\caption[short table name.]{
Resonance-to-bound state transition rates are displayed along with the corresponding x-ray energies. 
$v_\mathrm{r}$ denotes the vibrational quantum number of resonance states, and $v_\mathrm{b}$ denotes that of bound states.
The rates are listed in descending order of the resonance-to-bound state transition rates.
The notation $x[y]$ represents $x\times10^{y}$.
}
\label{tbl_rr-rb}
\begin{tabular}{rlrlrr}
\multicolumn{2}{c}{Initial state} & \multicolumn{2}{c}{Final state} & $E_\gamma$ (eV) &  $\varGamma_\mathrm{RB(RR)}$ (ps$^{-1}$) \\
\hline
$^{3}D^{e}$& $v_\mathrm{r}=0^\ast$   &  $^{3}P^{o}$& $v_\mathrm{b}=1$  &    $1989.35$ &  $5.624[-2]$ \\
$^{1,5}P^{o}$& $v_\mathrm{r}=0^\ast$ &  $^{1,5}S^{e}$& $v_\mathrm{b}=1$  &    $2010.56$ &  $3.513[-2]$ \\
$^{3}S^{e}$& $v_\mathrm{r}=0$        &  $^{3}P^{o}$& $v_\mathrm{b}=1$  &    $1978.18$ &  $5.658[-3]$ \\
$^{3}D^{e}$& $v_\mathrm{r}=0$        &  $^{3}P^{o}$& $v_\mathrm{b}=1$  &    $1981.19$ &  $3.442[-3]$ \\
$^{1,5}P^{o}$& $v_\mathrm{r}=0^\ast$ &  $^{1,5}D^{e}$& $v_\mathrm{b}=0$  &    $2061.21$ &  $3.301[-3]$ \\
$^{1,5}P^{o}$& $v_\mathrm{r}=0^\ast$ &  $^{1,5}S^{e}$& $v_\mathrm{b}=0$  &    $2299.78$ &  $9.314[-4]$ \\
$^{3}D^{e}$& $v_\mathrm{r}=0^\ast$   &  $^{3}P^{o}$& $v_\mathrm{b}=0$  &    $2214.05$ &  $7.683[-4]$ \\
$^{1,5}P^{o}$& $v_\mathrm{r}=0^\ast$ &  $^{1,5}S^{e}$& $v_\mathrm{r}=0$  &    $ 195.46$ &  $4.440[-4]$ \\
$^{3}S^{e}$& $v_\mathrm{r}=2$        &  $^{3}P^{o}$& $v_\mathrm{b}=1$  &    $1995.26$ &  $4.411[-4]$ \\
$^{3}D^{e}$& $v_\mathrm{r}=0^\ast$   &  $^{3}P^{o}$& $v_\mathrm{r}=0$    &    $ 201.93$ &  $3.711[-4]$ \\
$^{3}S^{e}$& $v_\mathrm{r}=1$        &  $^{3}P^{o}$& $v_\mathrm{b}=1$  &    $1989.92$ &  $3.660[-4]$ \\
$^{3}D^{e}$& $v_\mathrm{r}=0^\ast$   &  $^{3}F^{o}$& $v_\mathrm{r}=0$    &    $ 172.47$ &  $2.710[-4]$ \\
$^{3}D^{e}$& $v_\mathrm{r}=1$        &  $^{3}P^{o}$& $v_\mathrm{b}=1$  &    $1991.73$ &  $2.667[-4]$ \\
$^{1,5}P^{o}$& $v_\mathrm{r}=0^\ast$ &  $^{1,5}D^{e}$& $v_\mathrm{r}=0$  &    $ 177.20$ &  $2.269[-4]$ \\
$^{3}S^{e}$& $v_\mathrm{r}=3$        &  $^{3}P^{o}$& $v_\mathrm{b}=1$  &    $1997.57$ &  $1.565[-4]$ \\
$^{1,5}P^{o}$& $v_\mathrm{r}=0$      &  $^{1,5}S^{e}$& $v_\mathrm{b}=1$  &    $2013.09$ &  $9.255[-5]$ \\
$^{3}S^{e}$& $v_\mathrm{r}=4$        &  $^{3}P^{o}$& $v_\mathrm{b}=1$  &    $1998.57$ &  $6.767[-5]$ \\
$^{3}S^{e}$& $v_\mathrm{r}=0$        &  $^{3}P^{o}$& $v_\mathrm{r}=1$    &    $ 109.19$ &  $3.514[-5]$ \\
$^{3}S^{e}$& $v_\mathrm{r}=0$        &  $^{3}P^{o}$& $v_\mathrm{r}=2$    &    $  48.07$ &  $3.215[-5]$ \\
$^{3}S^{e}$& $v_\mathrm{r}=1$        &  $^{3}P^{o}$& $v_\mathrm{r}=1$    &    $ 120.93$ &  $3.022[-5]$ \\
$^{3}S^{e}$& $v_\mathrm{r}=5$        &  $^{3}P^{o}$& $v_\mathrm{b}=1$  &    $1999.01$ &  $2.923[-5]$ \\
$^{3}D^{e}$& $v_\mathrm{r}=2$        &  $^{3}P^{o}$& $v_\mathrm{b}=1$  &    $1996.19$ &  $2.870[-5]$ \\
$^{1,5}P^{o}$& $v_\mathrm{r}=0$      &  $^{1,5}D^{e}$& $v_\mathrm{r}=1$  &    $ 100.62$ &  $2.129[-5]$ \\
$^{3}S^{e}$& $v_\mathrm{r}=1$        &  $^{3}P^{o}$& $v_\mathrm{r}=2$    &    $  59.81$ &  $1.859[-5]$ \\
$^{1,5}P^{o}$& $v_\mathrm{r}=1$      &  $^{1,5}D^{e}$& $v_\mathrm{r}=1$  &    $ 111.94$ &  $1.842[-5]$ \\
$^{1,5}P^{o}$& $v_\mathrm{r}=0$      &  $^{1,5}D^{e}$& $v_\mathrm{b}=0$  &    $2063.73$ &  $1.713[-5]$ \\
$^{3}D^{e}$& $v_\mathrm{r}=0$        &  $^{3}F^{o}$& $v_\mathrm{r}=1$    &    $  88.83$ &  $1.703[-5]$ \\
$^{1,5}P^{o}$& $v_\mathrm{r}=0$      &  $^{1,5}D^{e}$& $v_\mathrm{r}=2$  &    $  41.90$ &  $1.639[-5]$ \\
$^{3}S^{e}$& $v_\mathrm{r}=2$        &  $^{3}P^{o}$& $v_\mathrm{r}=1$    &    $ 126.26$ &  $1.597[-5]$ \\
$^{3}D^{e}$& $v_\mathrm{r}=1$        &  $^{3}P^{o}$& $v_\mathrm{r}=0$    &    $ 204.31$ &  $1.480[-5]$ \\
$^{3}D^{e}$& $v_\mathrm{r}=1$        &  $^{3}P^{o}$& $v_\mathrm{b}=0$  &    $2216.43$ &  $1.394[-5]$ \\
$^{3}D^{e}$& $v_\mathrm{r}=3$        &  $^{3}P^{o}$& $v_\mathrm{b}=1$  &    $1998.04$ &  $1.382[-5]$ \\
$^{3}D^{e}$& $v_\mathrm{r}=0$        &  $^{3}P^{o}$& $v_\mathrm{r}=2$    &    $  51.09$ &  $1.313[-5]$ \\
$^{1,5}P^{o}$& $v_\mathrm{r}=0$      &  $^{1,5}S^{e}$& $v_\mathrm{r}=2$  &    $  52.84$ &  $1.235[-5]$ \\
$^{1,5}F^{o}$& $v_\mathrm{r}=0$      &  $^{1,5}G^{e}$& $v_\mathrm{r}=1$  &    $  72.90$ &  $1.155[-5]$ \\
$^{1,5}P^{o}$& $v_\mathrm{r}=0$      &  $^{1,5}S^{e}$& $v_\mathrm{r}=1$  &    $ 115.15$ &  $1.151[-5]$ \\
$^{1,5}F^{o}$& $v_\mathrm{r}=0$      &  $^{1,5}D^{e}$& $v_\mathrm{r}=2$  &    $  46.85$ &  $1.144[-5]$ \\
$^{3}D^{e}$& $v_\mathrm{r}=1$        &  $^{3}P^{o}$& $v_\mathrm{r}=1$    &    $ 122.74$ &  $1.119[-5]$ \\
$^{3}D^{e}$& $v_\mathrm{r}=1$        &  $^{3}F^{o}$& $v_\mathrm{r}=1$    &    $  99.37$ &  $1.118[-5]$ \\
$^{3}D^{e}$& $v_\mathrm{r}=0$        &  $^{3}P^{o}$& $v_\mathrm{r}=1$    &    $ 112.20$ &  $1.089[-5]$ 
\end{tabular}
\end{ruledtabular}
\end{table}

Table~\ref{tbl_ddmueng} summarizes the radiative decay rates for each resonance state.
The total decay rate into a continuum, denoted by $\varGamma_\mathrm{RC}$, 
is obtained by numerical integration of the x-ray spectrum, and
is a sum of 
$\varGamma_\mathrm{RC(-)}$ and $\varGamma_\mathrm{RC(+)}$, which are
the $J$-decreasing and $J$-increasing decay rates, respectively. 
The radiative decay rates listed in the table are obtained using a velocity gauge, and the uncertainties are estimated from the 
comparison with the length-gauge calculations.
For a few high vibrational states, 
the length-gauge calculation resulted in unphysical values for the radiative decay into the continuum and the rates.
As the bound states exist only in $^{1,5}S^e$, $^{3}P^o$, and $^{1,5}D^e$ symmetries, the resonance states $v_\mathrm{r}\leq3$
in $^{1,5}S^e$, $^{3}P^o$, $^{1,5}D^e$, and $^{3}F^o$ symmetries exhibited neither resonance-to-resonance nor resonance-to-bound 
decay branches under dipole approximation.
$\varGamma_\mathrm{RC}$ can be compared with a d$\murm(2p)\to$d$\murm(1s)$ transition rate of $0.122\,70$ ps$^{-1}$.
As shown in Fig.~\ref{f_dipole_ser}, the $\ddmr$ resonance states at high vibrational states can be considered as members of the 
dipole series in which the wave function involves a mixed fraction of d$\murm(2s)$ and d$\murm(2p)$.
The radiative decay rates for the high vibrational states are close to 
the half value of the d$\murm(2p)\to$d$\murm(1s)$ transition rate, $0.06135$ ps$^{-1}$.

\begin{table*}[p]                
\begin{ruledtabular}                
\caption[short table name.]{
Resonance energies, 
decay rates into continuum ($\varGamma_\mathrm{RC(-)}$, $\varGamma_\mathrm{RC(+)}$, and $\varGamma_\mathrm{RC}=\varGamma_\mathrm{RC(-)}+\varGamma_\mathrm{RC(+)}$) and
decay rates into other resonance or bound states ($\varGamma_\mathrm{RB+RR'}$) are listed.
The uncertainty is estimated by the difference between the velocity and length gauge calculations and are noted in parentheses. 
The decay rates with $\dag$ are obtained solely from the velocity-gauge calculation.
The notation $x[y]$ represents $x\times10^{y}$.
}
\label{tbl_ddmueng}
\begin{tabular}{cccrcccc}
Symmetry    &  $v_\mathrm{r}$ & $E_{J,v_\mathrm{r}}^{\mathrm{(R)}}$ (m.a.u.) & $\varepsilon_{J,v_\mathrm{r}}$ (eV) 
            & $\varGamma_\mathrm{RC(-)}$ (ps$^{-1}$) & $\varGamma_\mathrm{RC(+)}$ (ps$^{-1}$) & $\varGamma_\mathrm{RC}$ (ps$^{-1}$)
            & $\varGamma_\mathrm{RB}+\varGamma_\mathrm{RR'}$ (ps$^{-1}$)  \\ \hline
$^{1,5}S^e$ & $ 0$ & $-0.1570992$ & $218.1111$ & & $7.95[-2]$   &  $8.0(1)[-2]$  & --                          \\  
$^{1,5}S^e$ & $ 1$ & $-0.1423772$ & $135.2785$ & & $6.97[-2]$   &  $7.0(1)[-2]$  & --                          \\  
$^{1,5}S^e$ & $ 2$ & $-0.1313023$ & $ 72.9662$ & & $6.36[-2]$   &  $6.4(2)[-2]$  & --                          \\  
$^{1,5}S^e$ & $ 3$ & $-0.1240038$ & $ 31.9011$ & & $6.10[-2]$   &  $6.1(2)[-2]$  & --                          \\  
$^{1,5}S^e$ & $ 4$ & $-0.1205763$ & $ 12.6165$ & & $6.15[-2]$   &  $6.2(1)[-2]$  & $3.413(2)[-7]$   \\
$^{1,5}S^e$ & $ 5$ & $-0.1192779$ & $  5.3112$ & & $6.15[-2]$   &  $6.15(3)[-2]$  & $1.676(1)[-7]$   \\
$^{1,5}S^e$ & $ 6$ & $-0.1187383$ & $  2.2750$ & & $6.15[-2]$   &  $6.15(1)[-2]$  & $7.477(5)[-8]$    \\
$^{1,5}S^e$ & $ 7$ & $-0.1185083$ & $  0.9810$ & & $6.23[-2]$   &  $6.23(1)[-2]$  & $3.251(2)[-8]$   \\
$^{1,5}S^e$ & $ 8$ & $-0.1184093$ & $  0.4241$ & & $5.52[-2]$   &  $5.52^\dag[-2]$  & $1.405(1)[-8]$    \\
\hline
$^{  3}P^o$ & $ 0$ & $-0.1559995$ & $211.9236$ & $2.96[-2]$ & $4.87[-2]$ & $7.8(2)[-2]$   & --                         \\  
$^{  3}P^o$ & $ 1$ & $-0.1415010$ & $130.3486$ & $2.61[-2]$ & $4.29[-2]$ & $6.9(2)[-2]$   & --                         \\  
$^{  3}P^o$ & $ 2$ & $-0.1306392$ & $ 69.2351$ & $2.38[-2]$ & $3.95[-2]$ & $6.3(2)[-2]$   & --                         \\  
$^{  3}P^o$ & $ 3$ & $-0.1235815$ & $ 29.5255$ & $2.29[-2]$ & $3.82[-2]$ & $6.1(2)[-2]$   & --                         \\ 
$^{  3}P^o$ & $ 4$ & $-0.1203769$ & $ 11.4945$ & $2.29[-2]$ & $3.85[-2]$ & $6.13(4)[-2]$   & $3.540(1)[-7]$  \\ 
$^{  3}P^o$ & $ 5$ & $-0.1191823$ & $  4.7732$ & $2.23[-2]$ & $3.84[-2]$ & $6.07(5)[-2]$   & $1.736(1)[-7]$      \\ 
$^{  3}P^o$ & $ 6$ & $-0.1186922$ & $  2.0157$ & $2.05[-2]$ & $3.84[-2]$ & $5.88(4)[-2]$   & $7.696(1)[-8]$      \\ 
$^{  3}P^o$ & $ 7$ & $-0.1184862$ & $  0.8567$ & $1.70[-2]$ & $3.89[-2]$ & $5.6(3)[-2]$   & $3.321(1)[-8]$     \\ 
$^{  3}P^o$ & $ 8$ & $-0.1183988$ & $  0.3650$ & $1.31[-2]$ & $4.08[-2]$ & $5.39^\dag[-2]$   & $1.421(1)[-8]$    \\ 
\hline
$^{1,5}D^e$ & $ 0$ & $-0.1538540$ & $199.8521$ & $3.65[-2]$ & $3.97[-2]$ & $7.6(2)[-2]$  & --                   \\                        
$^{1,5}D^e$ & $ 1$ & $-0.1397954$ & $120.7520$ & $3.23[-2]$ & $3.51[-2]$ & $6.7(3)[-2]$  & --                   \\                        
$^{1,5}D^e$ & $ 2$ & $-0.1293582$ & $ 62.0277$ & $2.97[-2]$ & $3.27[-2]$ & $6.2(2)[-2]$  & --                   \\                        
$^{1,5}D^e$ & $ 3$ & $-0.1227887$ & $ 25.0647$ & $2.90[-2]$ & $3.20[-2]$ & $6.1(2)[-2]$  & --                   \\                        
$^{1,5}D^e$ & $ 4$ & $-0.1200140$ & $  9.4526$ & $2.95[-2]$ & $3.23[-2]$ & $6.2(1)[-2]$  & $3.31(3) [-7]$      \\
$^{1,5}D^e$ & $ 5$ & $-0.1190108$ & $  3.8085$ & $2.97[-2]$ & $3.22[-2]$ & $6.19(9)[-2]$  & $1.602(1)[-7]$     \\
$^{1,5}D^e$ & $ 6$ & $-0.1186110$ & $  1.5590$ & $2.98[-2]$ & $3.16[-2]$ & $6.2(2)[-2]$  & $6.92(2) [-8]$      \\
$^{1,5}D^e$ & $ 7$ & $-0.1184481$ & $  0.6421$ & $3.02[-2]$ & $2.98[-2]$ & $6.18^\dag[-2]$  & $2.90(1) [-8]$      \\
\hline
$^{  3}F^o$ & $ 0$ & $-0.1507635$ & $182.4632$ & $3.93[-2]$ & $3.37[-2]$ & $7.3(2)[-2]$  & --                 \\ 
$^{  3}F^o$ & $ 1$ & $-0.1373475$ & $106.9787$ & $3.50[-2]$ & $3.01[-2]$ & $6.5(4)[-2]$  & --                 \\ 
$^{  3}F^o$ & $ 2$ & $-0.1275472$ & $ 51.8380$ & $3.26[-2]$ & $2.86[-2]$ & $6.1(3)[-2]$  & --                 \\ 
$^{  3}F^o$ & $ 3$ & $-0.1217280$ & $ 19.0964$ & $3.24[-2]$ & $2.84[-2]$ & $6.1(2)[-2]$  & --                 \\ 
$^{  3}F^o$ & $ 4$ & $-0.1195513$ & $  6.8493$ & $3.35[-2]$ & $2.85[-2]$ & $6.20(1)[-2]$  & $2.240(1)[-7]$  \\ 
$^{  3}F^o$ & $ 5$ & $-0.1187990$ & $  2.6166$ & $3.37[-2]$ & $2.85[-2]$ & $6.2(3)[-2]$  & $1.025(1)[-7]$   \\ 
$^{  3}F^o$ & $ 6$ & $-0.1185143$ & $  1.0147$ & $3.42[-2]$ & $2.83[-2]$ & $6.24^\dag[-2]$  & $4.14(1) [-8]$  \\ 
$^{  3}F^o$ & $ 7$ & $-0.1184043$ & $  0.3957$ & $3.59[-2]$ & $2.78[-2]$ & $6.38^\dag[-2]$  & $1.640(1)[-8]$  \\ 
\hline
$^{  3}S^e$ & $ 0$ & $-0.1220939$ & $ 21.1551$ &  & $5.77[-2]$ & $5.77(7)[-2]$  & $5.733(1)[-3]$         \\
$^{  3}S^e$ & $ 1$ & $-0.1200073$ & $  9.4149$ &  & $6.18[-2]$ & $6.2(3)[-2]$  & $4.217(1)[-4]$         \\
$^{  3}S^e$ & $ 2$ & $-0.1190591$ & $  4.0801$ &  & $6.13[-2]$ & $6.13(6)[-2]$  & $4.692(1)[-4]$        \\
$^{  3}S^e$ & $ 3$ & $-0.1186477$ & $  1.7656$ &  & $6.13[-2]$ & $6.13(6)[-2]$  & $1.694(1)[-4]$         \\
$^{  3}S^e$ & $ 4$ & $-0.1184698$ & $  0.7645$ &  & $6.10[-2]$ & $6.10(7)[-2]$  & $7.337(1)[-5]$         \\
$^{  3}S^e$ & $ 5$ & $-0.1183928$ & $  0.3311$ &  & $4.59[-2]$ & $4.59^\dag[-2]$  & $3.173(1)[-5]$          \\
\hline
$^{1,5}P^o$ & $ 0^\ast$ & $-0.1223588$ & $ 22.6458$ & $4.83[-2]$ & $3.36[-2]$ & $8.2(3)[-2]$   & $4.004(1)[-2]$     \\
$^{1,5}P^o$ & $ 0$ & $-0.1219101$ & $ 20.1211$ & $2.37[-2]$ & $3.94[-2]$ & $6.3(2)[-2]$  & $1.854(1)[-4]$   \\
$^{1,5}P^o$ & $ 1$ & $-0.1198988$ & $  8.8046$ & $2.35[-2]$ & $3.85[-2]$ & $6.2(1)[-2]$  & $5.371(1)[-5]$   \\
$^{1,5}P^o$ & $ 2$ & $-0.1190018$ & $  3.7575$ & $2.21[-2]$ & $3.83[-2]$ & $6.04(8)[-2]$  & $2.714(1)[-5]$  \\
$^{1,5}P^o$ & $ 3$ & $-0.1186187$ & $  1.6023$ & $1.98[-2]$ & $3.83[-2]$ & $5.81(8)[-2]$  & $1.275(1)[-5]$  \\
$^{1,5}P^o$ & $ 4$ & $-0.1184554$ & $  0.6837$ & $1.55[-2]$ & $3.90[-2]$ & $5.45(7)[-2]$  & $5.780(1)[-6]$   \\
$^{1,5}P^o$ & $ 5$ & $-0.1183858$ & $  0.2918$ & $1.25[-2]$ & $3.92[-2]$ & $5.17^\dag[-2]$  & $2.557(1)[-6]$   \\
\hline
$^{  3}D^e$ & $ 0$ & $-0.1215586$ & $ 18.1432$ & $2.86[-2]$ & $3.16[-2]$ & $6.0(2)[-2]$  & $3.501(1)[-3]$      \\
$^{  3}D^e$ & $ 0^\ast$ & $-0.1201091$ & $  9.9880$ & $1.83[-2]$ & $4.03[-2]$ & $5.9(3)[-2]$  & $5.766(1)[-2]$     \\
$^{  3}D^e$ & $ 1$ & $-0.1196858$ & $  7.6061$ & $2.92[-2]$ & $3.36[-2]$ & $6.3(2)[-2]$  & $3.306(1)[-4]$    \\
$^{  3}D^e$ & $ 2$ & $-0.1188932$ & $  3.1468$ & $2.97[-2]$ & $3.23[-2]$ & $6.21(5)[-2]$  & $4.979(1)[-5]$    \\
$^{  3}D^e$ & $ 3$ & $-0.1185649$ & $  1.2993$ & $2.98[-2]$ & $3.21[-2]$ & $6.19(3)[-2]$  & $2.276(1)[-5]$    \\
$^{  3}D^e$ & $ 4$ & $-0.1184294$ & $  0.5369$ & $2.89[-2]$ & $3.20[-2]$ & $6.09(6)[-2]$  & $1.016(1)[-5]$    \\
\hline
$^{1,5}F^o$ & $ 0$ & $-0.1210306$ & $ 15.1726$ & $3.60[-2]$ & $2.76[-2]$ & $6.4(1)[-2]$  & $4.2(1)  [-5]$     \\
$^{1,5}F^o$ & $ 1$ & $-0.1194050$ & $  6.0264$ & $3.50[-2]$ & $2.80[-2]$ & $6.3(1)[-2]$  & $2.803(5)[-5]$     \\
$^{1,5}F^o$ & $ 2$ & $-0.1187524$ & $  2.3547$ & $3.46[-2]$ & $2.80[-2]$ & $6.27(8)[-2]$  & $1.357(7)[-5]$    \\
$^{1,5}F^o$ & $ 3$ & $-0.1184976$ & $  0.9210$ & $3.46[-2]$ & $2.81[-2]$ & $6.3(1)[-2]$  & $6.07(5) [-6]$     \\
$^{1,5}F^o$ & $ 4$ & $-0.1183980$ & $  0.3604$ & $3.55[-2]$ & $2.83[-2]$ & $6.38(9)[-2]$  & $2.55(3) [-6]$    
\end{tabular}
\end{ruledtabular}
\end{table*}

\subsection{Branching ratios into bound state from $\mathbf{dd}\murm^\ast$ and $\mathbf{dt}\murm^\ast$}

\begin{figure*}[tb]
\begin{center}
\resizebox{0.95\textwidth}{!}{%
\includegraphics[bb=0 0 680 283]{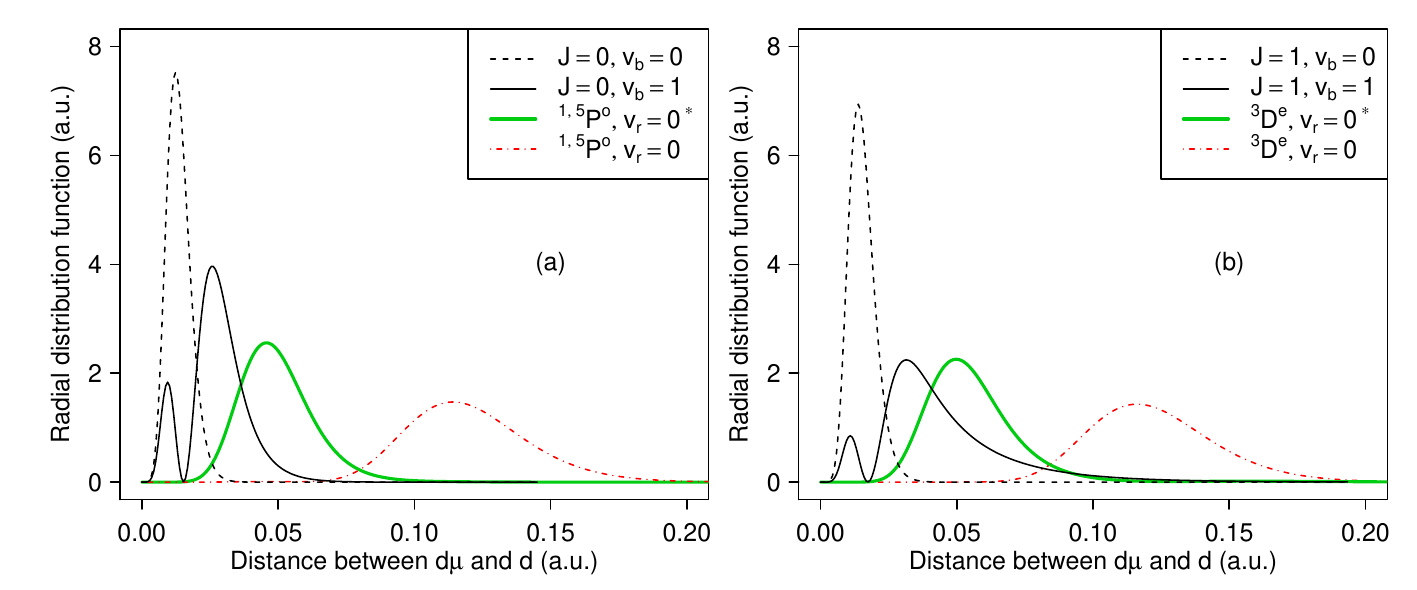}}%
\caption{
\add{
Radial distribution functions of bound and resonance states of $\ddmr$ are shown as a function of distance between d$\murm$ and d. 
(a) The $^{1,5}P^o$ resonance states ($v_\mathrm{r}=0, 0^\ast$) are compared with 
the $^{1,5}S^e$ bound states, $(J,v_\mathrm{b})=(0,0)$ and $(0,1)$. 
(b) The $^{3}D^e$ resonance states ($v_\mathrm{r}=0, 0^\ast$) are compared with 
the $^{3}P^o$ bound states, $(J,v_\mathrm{b})=(1,0)$ and $(1,1)$. 
}
}
\label{f_corr_overlap}
\end{center}
\end{figure*}

It has been unexpected that the radiative decay rates into the bound and resonance states ($\varGamma_\mathrm{RB+RR'}$) of $v_\mathrm{r}=0^\ast$ states in $^{3}D^{e}$ and $^{1,5}P^{o}$
are comparable with the radiative decay rates into the continuum.
\add{
The high $\varGamma_\mathrm{RB+RR'}$ is mainly contributed from the radiative decay rate into the bound states, $\varGamma_\mathrm{RB}$.
In Fig.~\ref{f_corr_overlap}, 
we compare the d$\murm$-d radial distribution functions 
of the resonance states and bound states.
As shown in Fig.~\ref{f_corr_overlap}(a), the radial distribution function of $^{1,5}P^o$ ($v_\mathrm{r}=0^\ast$) has an overlap
with the $J_\mathrm{f}=0,v_\mathrm{b}=1$ bound state. 
In the similar way, as shown in Fig.~\ref{f_corr_overlap}(b), the $^{3}D^e$ ($v_\mathrm{r}=0^\ast$) resonance state
has an overlap in the radial distribution function with the $J_\mathrm{f}=1,v_\mathrm{b}=1$ bound state.
It is also noticeable that the $J_\mathrm{f}=1,v_\mathrm{b}=1$ bound state has a long-range tail which has a small overlap 
with the radial distribution function of $^{3}D^e$ ($v_\mathrm{r}=0$) resonance state. 
These observations are consistent with the fact that the resonance states $v_\mathrm{r}=0^\ast$ of $^{1,5}P^o$ and $^{3}D^e$
show relatively high $\varGamma_\mathrm{RB}$.
}

The branching ratio of the decay rate into the bound state ($\varGamma_\mathrm{RB}$) against the total radiative decay rate ($\varGamma_\mathrm{RC}+\varGamma_\mathrm{RB}$)
\begin{align}
\Upsilon_\mathrm{RB}= \frac{\varGamma_\mathrm{RB}}{\varGamma_\mathrm{RC}+\varGamma_\mathrm{RB}},
\end{align}
is an intriguing quantity for applications of \mcf. 
The radiative decay into the bound state might be a fast track in the $\murm$CF cycle because it skips the slow process of the bound state formation 
and accelerate the \mcf\ cycles.
Moreover, the decay into the continuum could prevent the muonic molecule formation and decelerate \mcf\ cycles. 
Therefore, we have calculated the radiative decay rates of $\dtmr$ in addition to the $\ddmr$ and examined the branching ratio $\Upsilon_\mathrm{RB}$ in Table~\ref{tbl_rb_ddm}.

\begin{table*}[t]                
\begin{ruledtabular}                
\caption[short table name.]{
Several states of $\ddmr$ and $\dtmr$ that have a high branching ratio resulting in a bound state muonic molecule. 
\add{Binding energies of $\ddmr$ and $\dtmr$ are measured from d$\murm(n=2)$ + d and t$\murm(n=2)$ + d threshold energies, respectively.}
The total angular momentum and the vibrational quantum number of the bound state
are denoted as $J_\mathrm{f}$ and $v_\mathrm{b}$, respectively.
$E_\gamma$ denotes the monoenergetic x-ray energy.
}
\label{tbl_rb_ddm}
  \begin{tabular}{ccccccccccc}
    Symmetry &  $v_\mathrm{r}$ & $E_{J,v_\mathrm{r}}^{\mathrm{(R)}}$(m.a.u.) & $\varepsilon_{J,v_\mathrm{r}}$ (eV) &  $\varGamma_\mathrm{RC}$ (ps$^{-1}$) & $\varGamma_\mathrm{Coul}$ (ps$^{-1}$) &  $J_\mathrm{f}$ & $v_\mathrm{b}$  &  $E_\gamma$ (eV)  & $\varGamma_\mathrm{RB}$ (ps$^{-1}$) &  $\Upsilon_\mathrm{RB}$ \\
    \hline
                 &          &              &            &  $\ddmr$    &   &   &    &            &            &           \\
     $^{  3}D^e$ & 0$^\ast$ & $-0.120\,109$ & $  9.9880$ & $5.9[-2]$ & $2.4[-3]$ & 1 &  1 &  $1989.35$ & $5.62[-2]$ & $0.49$    \\
     $^{1,5}P^o$ & 0$^\ast$ & $-0.122\,359$ & $ 22.6458$ & $8.2[-2]$ & $3.5[-3]$ & 0 &  1 &  $2010.56$ & $3.51[-2]$ & $0.30$    \\
     $^{  3}S^e$ & 0        & $-0.122\,094$ & $ 21.1551$ & $5.8[-2]$ & $(<1[-3])$ & 1 &  1 &  $1978.18$ & $5.66[-3]$ & $0.09$    \\
     $^{  3}D^e$ & 0        & $-0.121\,559$ & $ 18.1432$ & $6.0[-2]$ & $(<8[-4])$ & 1 &  1 &  $1981.19$ & $3.44[-3]$ & $0.05$    \\  
                 &          &              &            &          &  &   &    &            &            &           \\
                 &          &              &            &  $\dtmr$ &  &   &    &            &            &           \\
             $D$ &  7       & $-0.121\,883$  & $7.957$   & $7.6[-2]$ & $8.9[-4]$ & 1  &  1  & $2026.72$ &  $2.68[-2]$  &  $0.35$ \\
             $P$ &  4       & $-0.123\,874$  & $19.159$  & $9.5[-2]$ & $1.2[-3]$ & 0  &  1  & $2049.10$ &  $2.50[-2]$  &  $0.26$ \\
             $S$ &  0       & $-0.159\,195$  & $217.887$ & $7.1[-2]$ & $2.9[-5]$\footnotemark & 1  &  1  & $1816.19$ &  $3.01[-3]$  &  $0.04$ \\ 
             $D$ &  1       & $-0.156\,357$  & $202.083$ & $7.4[-2]$ & $(<1[-3])$ & 1  &  1  & $1832.16$ &  $2.79[-3]$  &  $0.04$ \\
             $D$ &  5       & $-0.122\,998$  & $14.272$  & $6.1[-2]$ & $(<1[-2])$ & 1  &  1  & $2019.85$ &  $1.39[-3]$  &  $0.02$ 
  \end{tabular}
\footnotetext[1]{Ref.~\cite{Kilic2004}}
\end{ruledtabular}
\end{table*}

In contrast to $\ddmr$, $\dtmr$ is a hetero-nuclear system, and the resonance series are separated only by the total angular momentum $J$.
The vibrational quantum number $v_\mathrm{r}$ of $\dtmr$ is determined from the lowest resonance state $v_\mathrm{r}=0$
although each resonance state has a major association with one of the three adiabatic series $2p\pi_u$, $3d\sigma_g$, and $4f\sigma_u$
and with one of the dissociation thresholds t$\murm(n=2)+$d and d$\murm(n=2)+$t.
As expected, a significantly high branching ratio $\Upsilon_\mathrm{RB}$ are obtained for the $D$ state ($J=2,v_\mathrm{r}=7$)
and $P$ state ($J=1,v_\mathrm{r}=4$). These resonance states are similar to the $v_\mathrm{r}=0^\ast$ resonance states of $^{3}D^e$
and $^{1,5}P^o$ in $\ddmr$ and can be categorized into the $2p\pi_u$ series.
The $J=2,v_\mathrm{r}=7$ resonance state results in $J_\mathrm{f}=1,v_\mathrm{b}=1$ bound state, 
which is a loosely bound state with a binding energy of only 0.66 eV.
The $J=1,v_\mathrm{r}=4$ state results in $J_\mathrm{f}=0,v_\mathrm{b}=1$, which has a binding energy of 34.8 eV.
Although the $\Upsilon_\mathrm{RB}$ for the other resonance states is less than those of these two states by one order of magnitude, 
the transition that results in a shallow-bound state ($J_\mathrm{f}=1,v_\mathrm{b}=1$) tends to exhibit relatively high transition rates.

\add{To conclude the significance of the radiative decay processes, 
we also calculate the non-radiative (Coulombic) decay rates $\varGamma_{\rm Coul}$ of these
resonance states by complex coordinate rotation trajectories and include them in Table~\ref{tbl_rb_ddm}. 
The resonance poles are clearly determined for resonance states of 
$^{3}D^e$ ($v_\mathrm{r}=0^\ast$) and $^{1,5}P^o$ ($v_\mathrm{r}=0^\ast$) of $\ddmr$
by the complex coordinate rotation trajectories.
For $\dtmr$, the poles of resonance states of $D$ ($v_\mathrm{r}=7$) and $P$ ($v_\mathrm{r}=4$)
are clearly determined by the complex coordinate rotation trajectories.
The resonance widths of $^{  3}S^e$ ($v_\mathrm{r}=0$), $^{  3}D^e$ ($v_\mathrm{r}=0$) of $\ddmr$
and $D$ ($v_\mathrm{r}=1,5$) of $\dtmr$ are estimated by the 3 times of strandard deviation of the trajectories near the real axis.  
The $\varGamma_{\rm Coul}$ for these resonance states are smaller than the $\varGamma_\mathrm{RC}+\varGamma_\mathrm{RB}$.
}

\begin{figure*}[bt]
\begin{center}
\resizebox{0.98\textwidth}{!}{%
\includegraphics[bb=0 0 1808 973]{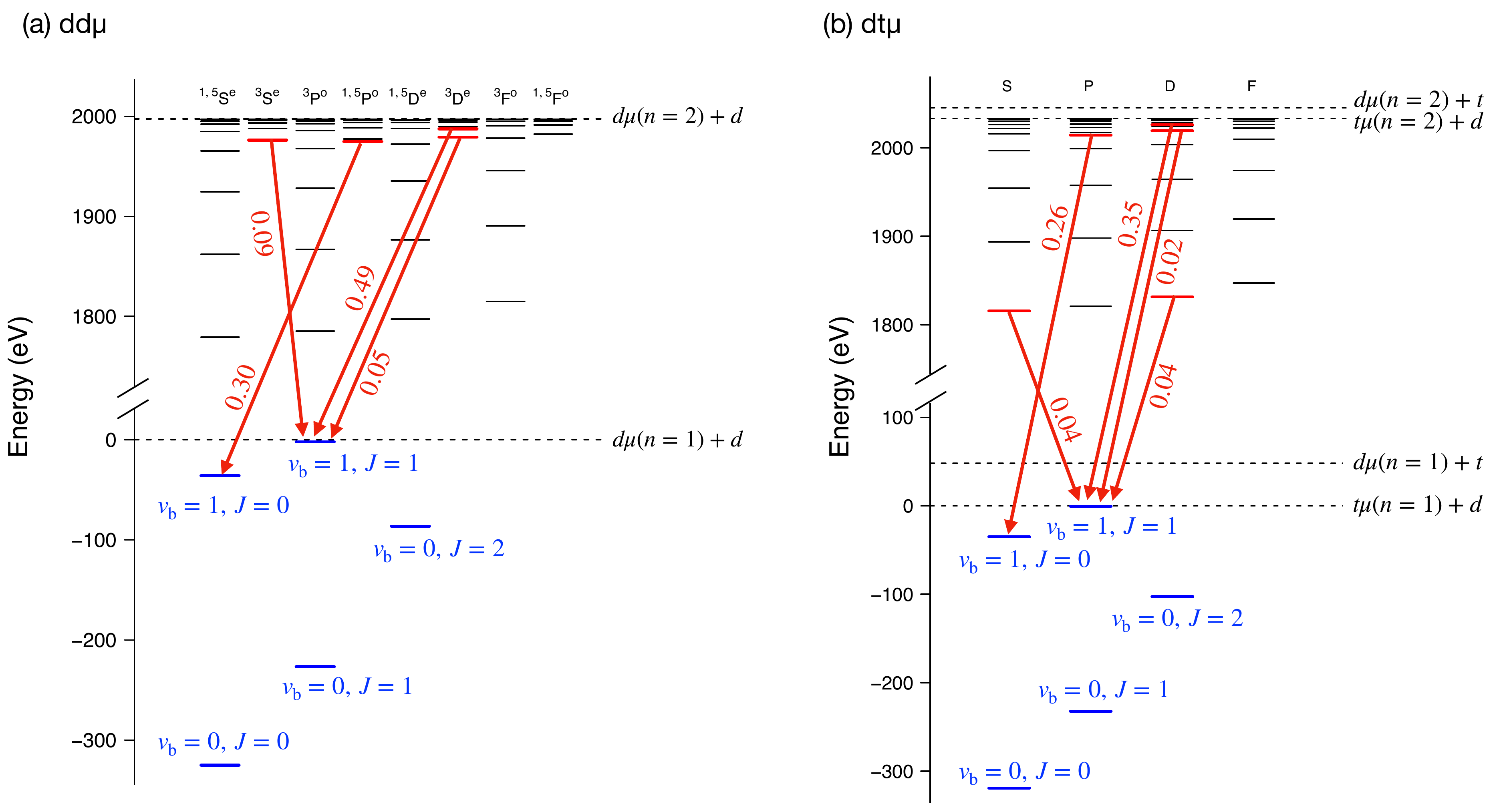}}%
\caption{
Energy diagram of resonance and bound energy levels for (a) $\ddm$ and (b) $\dtm$. The red arrows highlight the transitions with a high branching ratio into the
bound state.
}
\label{f_reso-to-bound}
\end{center}
\end{figure*}

Figure~\ref{f_reso-to-bound} illustrates the resonance-to-bound transition lines 
with the high branching ratio into the bound state $\Upsilon_\mathrm{RB}$.
The resonance states exhibiting high $\Upsilon_\mathrm{RB}$ values have binding energies of over a few eV 
that is outside the energy range where the Vesman mechanism of Eq.~(\ref{vesman}) occurs.
However, the $^{1,5}P^o$ ($v_\mathrm{r}=0^\ast$) states of $\ddmr$ and the $P$($v_\mathrm{r}=4$) state of $\dtmr$ have a binding energy $\varepsilon_{J,v_\mathrm{r}}$
greater than 15.4 eV, i.e., the Auger transitions could produce them from the shallower resonance states formed by the Vesman mechanism.
Future research should comprise an investigation into the target conditions and laser-assisted processes 
that efficiently produce these resonance states.

\section{Conclusion}
\label{conclusion}

We comprehensively studied the x-ray spectra from radiative decay of dd$\murm^\ast$
that play a central role in the new kinetic model of $\murm$CF. 
The x-ray spectra predicted in this study can be used when they are measured in a future precise x-ray spectroscopy experiment 
and provide a fundamental understanding of muonic molecular dynamics.

Using the complex coordinate rotation method, the x-ray spectra from the decay into a continuum were calculated.
We determined the characteristic shapes of the spectra that depend on the rovibrational states of the resonance states of dd$\murm$.
The shapes of the x-ray spectra are examined for complex rotation angles, number of complex-rotated wave functions, and length and velocity gauges.
We pointed out that the centrifugal potential in the final state blurs the shape of x-ray spectrum.

The calculated x-ray spectrum provides the kinetic energy distribution of the decay fragments which shows that the 
radiative decay of dd$\murm^\ast$ produces muonic atoms with several tens eV.
The kinetic energy distribution deviates from that derived from the adiabatic approximation, indicating the importance of the non-adiabatic treatment.
The radiative decay can be a source of epi-thermal muonic atoms in $\murm$CF cycle.

We also calculated the radiative decay rates into the bound state, 
and compared them with the radiative decay rates into the continuum.
We found that some states of dd$\murm^\ast$ and dt$\murm^\ast$ demonstrate significantly high branching ratios into the bound state,
which can be a fast track in $\murm$CF 
because the efficient formation of dt$\murm^\ast$ and subsequent radiative decay can skip rate-limiting processes of $\murm$CF cycle, 
namely, the muon transfer from d$\murm(1s)$ to triton and muonic molecule formation via the Vesman mechanism.
Since the formation of the $\dtmr$ is based on the energy matching between the formation energy and excitation energy of the D$_2$,
the formation of the $\dtmr$ related to the fast track could be induced by changing the temperature and population of rovibrational level distribution of D$_2$.

\begin{acknowledgments}
We are grateful to S. Okada and Y. Toyama (Chubu University, Japan) for information about the feasibility of detecting the x-ray 
spectra using a TES microcalorimeter.
T.Y. is thankful for the 
financial support received from the Japan Society for the Promotion of
Science (JSPS) KAKENHI Grant No. JP24K06911. 
Y.K. is thankful for the JSPS KAKENHI Grant No. JP24K00549.

This work was partly achieved using the supercomputer systems at Hokkaido University and Kyushu University.
\end{acknowledgments}

\end{document}